\documentclass[a4paper]{elsarticle}

\usepackage{lipsum}
\usepackage[utf8]{inputenc}
\usepackage{import}
\usepackage{comment}

\usepackage{fontawesome}

\usepackage{fullpage}
\usepackage{geometry}
\usepackage{circledsteps}
\pgfkeys{/csteps/fill color=white}
\pgfkeys{/csteps/inner ysep=5pt}
\pgfkeys{/csteps/inner xsep=5pt}

\usepackage[english]{babel}

\usepackage{enumitem}
\setlist[1]{itemsep=-5pt}

\usepackage{amsmath,amssymb,amsthm}
\usepackage{cases}
\usepackage{mathrsfs}
\usepackage{bm}

\usepackage{algorithm}
\usepackage{algpseudocode}
\theoremstyle{definition}

\usepackage{stmaryrd}
\usepackage{cases} 
\usepackage{appendix}
\providecommand{\appendixname}{Appendix}
\usepackage{pdflscape}
\usepackage{tabularx}
\usepackage{booktabs}         

\usepackage{multicol}
\usepackage{multirow}
\usepackage{xparse}
\usepackage{xspace}
\usepackage{makecell}

\usepackage{xcolor}
\usepackage{footnote} 
\usepackage{graphicx}
\usepackage{caption,subcaption,tikz}
\usepackage{svg}
\usepackage{epstopdf}

\definecolor{ultramarine}{RGB}{0,32,96}
\definecolor{fig1Bt}{RGB}{0,11,178}
\definecolor{fig1Bp}{RGB}{100,100,100}
\definecolor{fig1bt}{RGB}{0,11,178}
\definecolor{fig1bp}{RGB}{100,100,100}

\usepackage{tikz}
\usetikzlibrary{calc}
\usepackage{pgfplots}
\pgfplotsset{compat=newest}

\usepackage[percent]{overpic}
\usepackage{placeins} 
\usepackage[colorlinks=true]{hyperref}         
\usepackage{cleveref}
\biboptions{sort&compress}
\usepackage[numbers]{natbib}
\usepackage{caption}          
\usepackage[notref,notcite,final]{showkeys}

\usepackage[color=gray, backgroundcolor=yellow, textwidth=2cm, textsize=footnotesize]{todonotes}
\usepackage{setspace}
\usepackage{subcaption}

\usepackage{framed} 
\usepackage[noprefix]{nomencl} 

\makenomenclature

\renewcommand*\nompreamble{\begin{multicols}{2}}
\renewcommand*\nompostamble{\end{multicols}}

\usepackage{etoolbox}
\usepackage{mathtools}

\usepackage{pdfpages}

\DeclareCaptionFormat{suboverlay}{\gdef\subcapoverlay{(\thesubfigure) #3\par}}
\DeclareCaptionStyle{suboverlay}{format=suboverlay}
\definecolor{DarkerGreen}{RGB}{0,111,0}
\definecolor{DarkerRed}{RGB}{156,0,0}
\definecolor{DarkerBlue}{RGB}{0,0,156}

\definecolor{myRed}{rgb}{0.8, 0.157961, 0.217975}

\definecolor{myKcolor}{rgb}{0, 0.411765, 0.572549}
\definecolor{myEcolor}{rgb}{0, 0, 0}

\newdefinition{rem}{Remark}

\DeclareMathAlphabet{\mathbb}{U}{bbold}{m}{n}

\makeatletter
\newcommand{\multiline}[1]{%
  \begin{tabularx}{\dimexpr\linewidth-\ALG@thistlm}[t]{@{}X@{}}
    #1
  \end{tabularx}
}
\makeatother

\newcommand{\dashedvert}{\tikz[baseline] \draw[dash pattern=on 1pt off 1pt] (0,-0.2em) -- (0,0.8em);}

\begin{document}

\begin{frontmatter}

\title{
Tensegrity structures and data-driven analysis for 3D cell mechanics
}


\author[aff1]{Ziran Zhou} 
\ead{zzhou2@caltech.edu}
\author[aff1,aff2]{Jacinto Ulloa}
\ead{julloa@umich.edu}
\author[aff1]{Guruswami Ravichandran}
\ead{ravi@caltech.edu}
\author[aff1]{José E. Andrade\corref{cor1}}
\ead{jandrade@caltech.edu}

\cortext[cor1]{Corresponding author}

 \address[aff1]{Division of Engineering and Applied Science, California Institute of Technology, Pasadena, CA 91125, USA}

\address[aff2]{Department of Mechanical Engineering, University of Michigan, Ann Arbor, MI 48109, USA}


\begin{abstract}
The cytoskeleton (CSK) plays an important role in many cell functions. Given the similarities between the mechanical behavior of tensegrity structures and the CSK, many studies have proposed different tensegrity-based models for simulating cell mechanics. However, the low symmetry of most tensegrity units has hindered the analysis of realistic 3D structures. As a result, tensegrity-based modeling in cell mechanics has been mainly focused on single cells or monolayers. In this paper, we propose a 3D tensegrity model based on the finite element method for simulating 3D cell mechanics. We show that the proposed model not only captures the nonlinearity of a single cell in an indentation test and a monolayer in stretch test but also the non-uniform stress distribution in multicellular spheroids upon non-uniform prestress design. Furthermore, we introduce a multiscale data-driven (DD) framework for cellular mechanics to optimize the computation, thus paving the way for modeling the mechanobiology of large cellular assemblies such as organs. 
\end{abstract}


\end{frontmatter}



\medskip

\section{Introduction}
\label{sec_intro}
\FloatBarrier

Living cells constantly experience complex interactions of mechanical forces \cite{Xi2019}. The ability of cells to sense and respond to such mechanical stimuli along with other biochemical cues is critical to cellular functionality \cite{Efremov2021,Zhu2000,Gomez2020}. These functions are mediated through changes in the cytoskeleton (CSK), which provides a structural basis for the mechanical and morphological behavior of cells \cite{wang1993_mechanotransdction_csk,SUN2023_feng, Hohmann2019,Ingber2014,Ingber2023}. The CSK comprises a dynamic network of proteins, mainly consisting of microtubules, intermediate filaments, and microfilaments. Through receptor proteins, the CSK senses and transmits forces to interact with the adjacent cells and microenvironment, in turn orchestrating cell shape changes, polarity, and motility \cite{Bashirzadeh2019}. An equilibrium in forces is maintained at the cellular level, balancing internal mechanical forces exerted by the contractile actin filaments, resisting microtubules in compression, and external forces from the extracellular matrix (ECM) \cite{Wendling2003}. This state of pre-existing stress in the CSK is referred to as prestress. Multiple studies have reported the central role of prestress in cellular mechanotransduction and mechanobiology \cite{Chowdhury2021, Hu2003,Huang1998,STAMENOVIC2005}. Altogether, changes to the mechanical properties of the CSK can lead to both cell development and disease \cite{Fletcher2010}.

Traditional computational efforts for simulating cell mechanics using the continuum approach describe the cell as an elastic, viscous, or viscoelastic medium, assuming the cell is composed of materials with certain continuum material properties. For example, the cortical shell–liquid core model assumes single or several layers of cortex with surface tension, with the cell interior modeled as Newtonian fluid with certain viscosity \cite{Yeung1989,Kan1998}. Other continuum models simulate the whole cell as a homogeneous solid \cite{Mijailovich2002,Schmid1981}. For a complete review, readers are referred to \cite{LIM2006}. Such continuum models are good approximations for a range of experimental observations. However, as experimental conditions become complex, constitutive models require additional parameters that are essentially empirical and often lack physical meaning. Moreover, it is difficult to consider subcellular processes, such as protein level activities, into such frameworks, except in an averaged sense \cite{FLETCHER2013}.

In individual cell-based approaches, most models can be grouped into two subsets: lattice-based models and off-lattice models \cite{Metzcar2019}. The former tracks cells along rigid grids, with examples being the fixed-lattice Cellular Automata and the Cellular Potts model \cite{FLETCHER2013}. The latter has no grid restrictions and the cells are tracked via centers of mass, as in the off-lattice center dynamics model \cite{Liedekerke2015}, or their boundaries, as in the vertex dynamics models \cite{FLETCHER2014}. In vertex dynamics models, additional biochemical rules can be employed to study more advanced cell behaviors such as cell rearrangements and oscillations \cite{FLETCHER2013,LIN2018}.

Most of the aforementioned modeling frameworks either cannot account for cellular prestress, despite its crucial role in determining cell behaviors, or lack physical insight in doing so. Models that account for the cell's internal structure or the CSK are still relatively scarce \cite{Liedekerke2015}. Moreover, it is difficult for the aforementioned models to directly link mechanical forces to specific load-bearing structures and molecules inside living cells, although such links are rather ubiquitous. For instance, Harris et al.~\cite{Harris2012} reported that depolymerizing actin CSK can result in a decrease in the stiffness of a monolayer by 50\%. It is therefore crucial for a model to incorporate not only prestress but also subcellular processes, such as distributing changes to specific groups of cell members, in order to further understand the mechanics of cells.

Complementing these frameworks, computational models based on micro- and nanostructural approaches offer a different perspective, focusing on the deformation of the CSK \cite{Pegoraro2017,STAMENOVIC2002, Wendling2003}. Among these, one category embraces the concept of tensegrity (tensional integrity), a structural principle originated in architecture by Buckminster Fuller \cite{Fuller1961}. Tensegrity is a self-stable assembly consisting of continuous tensile elements (tendons) and discontinuous compressive elements (bars). Tendons and bars are pin-connected in force equilibrium, setting the tensegrity structure in a tensile prestressed state. 

The tensegrity principle is observed in various natural systems across scales, from the kinematics of human and animal locomotion involving bones and tendons, down to the molecular architecture of spider silk fibers \cite{skelton2009tensegrity}. Within the CSK, the force balance between contractile actin filaments and compression-supporting structures, such as microtubules \cite{Ingber2014}, is reminiscent of the self-equilibrium of tensegrity structures. Moreover, studies have shown that cell stiffness scales with the CSK prestress, in accordance with predictions from tensegrity models \cite{Wang2001,wang2001_2}. These factors render tensegrity a conceptually simple but relevant idea for simulating cell~mechanics.

In this context, Sun et al.~\cite{SUN2023_feng} combined the CSK-based tensegrity model with a biochemical mechanism to simulate more realistic cellular-level processes. Their study, however, is largely focused on the biochemomechanical processes of a single cell. The collective response of tensegrity structures as an assembly of multiple cells is not investigated. Liu et al.~\cite{LIU2021} proposed a tensegrity model combined with the vertex dynamics method to explain the stiffness difference between a single cell and a monolayer. Moreover, Wang et al.~\cite{Wang2022_XuGK} proposed a multiscale model to investigate the static and dynamic response of cell monolayers. Due to the presence of torsion components in most tensegrity structures, application of multicellular tensegrity in cell mechanics has mostly been limited to 2D \cite{RIMOLI2017}; see, for instance, \cite{LIU2021} and \cite{Wang2022_XuGK}. 

In this paper, we propose a 3D multicellular tensegrity model capable of simulating cells in one-, two-, and three-dimensions, and investigate the response of such structure under different loading conditions. The tensegrity modeling framework is based on the finite element method (FEM) in the finite (large) deformation setting and allows controlling initial prestress \cite{MA2019a,MA2022}. As with any FEM model, the computational cost becomes exceedingly expensive as the number of elements increases, especially for our proposed use in cell mechanics. Therefore, to reduce the computational cost, we further perform simulations using a multiscale data-driven (DD) approach in concert with homogenization techniques. Data-driven computing enables calculations directly from a material dataset while satisfying pertinent constraints and conservation laws, thus bypassing the empirical material modeling step in the conventional FEM \cite{KIRCHDOERFER2016}. The material (training) dataset can be extracted from lower-scale computations or experimental observations.

The structure of this paper is as follows. In Section~\ref{sec:tenseg} and~\ref{sec:cell}, we present a brief overview of the tensegrity formulation, followed by simulations of a single cell, a monolayer, and a multicellular spheroid, each compared with published experimental work. In Section \ref{sec:DD}, we briefly outline the DD framework, followed by3D monolayer simulations, employing material data extracted from representative volumes of tensegrity structures. The DD solutions are then benchmarked against the tensegrity-based direct numerical simulations.

\section{Tensegrity structures}
\label{sec:tenseg}

\subsection{Modeling framework}
A tensegrity structure is composed of pin-connected tensile and compressive elements. In the present work, we mostly adopt the modeling framework of Ma et al.~\cite{MA2022}, briefly summarized below.

Representing the Cartesian coordinates of each node as $\bm n_i = [x_i\ y_i\ z_i]^\mathrm{T}$, with $i = 1,2,\dots,N$ indexing the node number, the global nodal coordinate matrix $\bm{N}\in\mathbb{R}^{3 \bigtimes N }$ can be constructed as
\begin{equation} 
    \bm N = [\bm n_1  \ \bm n_2 \ \dots \  \bm n_{N}]\,.
\end{equation}
In a single-vector formulation, the nodal coordinate matrix reads 
\begin{equation} 
    \bm n = [\bm n_1^\mathrm{T} \ \bm n_2^\mathrm{T} \ \dots \ \bm n_{N}^\mathrm{T}]^\mathrm{T}\,.
\end{equation}

The connectivity matrix $\bm C \in \mathbb{R}^{M \bigtimes N}$ represents the topology of the entire tensegrity structure, which consists of $M_{\mathrm{b}}$ compressive bars and $M_{\mathrm{s}}$ tensile strings. For each element $e_{ij}^k$ connecting node $\bm n_i$ and $\bm n_j$, with $k = 1,2,\dots,M$,  $M = M_{\mathrm{b}} + M_{\mathrm{s}}$, its corresponding $k$th row $l$th column entry $C_{kl}$ satisfies
\begin{equation}
    C_{kl} = \left\{ \begin{array}{cl}
                    1 &  \ \mathrm{if} \ l = i \,,\\
                    -1 &  \ \mathrm{if} \ l = j \,,\\
                    0 & \ \mathrm{otherwise}\,.
                    \end{array} \right.
\end{equation}
Each element $e^k_{ij}$ has a length $l_k$, calculated as 
\begin{equation} 
    l_k = \lVert \bm n_i - \bm n_j  \rVert\,.
\end{equation}
Accordingly, the global structure length vector $\bm l \in\mathbb{R}^{M}$ is assembled as
\begin{equation} 
    \bm l = \big[l_1 \ l_2 \ \dots \  l_{M}\big]^\mathrm{T}\,.
\end{equation}
Similarly, we define the global area vector $\bm A \in\mathbb{R}^{M}$ and modulus vector $\bm E \in\mathbb{R}^{M}$, given by
\begin{equation} 
    \bm A = [A_1 \ A_2 \ \dots \  A_{M}]^\mathrm{T}\,,
\end{equation}
\begin{equation}
    \bm E = [E_{1} \ E_{2} \ \dots \  E_{M}]^\mathrm{T}\,.
\end{equation}

In each element $k$, the (axial) stress increment $\mathrm{d}\sigma_k$ is given in terms of the (axial) strain increment $\mathrm{d}\epsilon_k$,
\begin{equation}
    \mathrm{d}\sigma_k = E_k \mathrm{d}\epsilon_k\,.
\end{equation}
The general tensegrity framework is equipped to handle elastic or plastic materials. In this study, we assume the members are elastic and have constant cross-sections and moduli, focusing on structural behavior under moderate strains without introducing additional complexity from nonlinear material behavior. Hence, the internal force vector $\bm t \in\mathbb{R}^{M}$ is assembled as
\begin{equation}
    \bm t = [t_{1} \ t_{2} \ \dots \  t_{M}] ^\mathrm{T} 
          = \mathrm{diag}(\bm E)\ \mathrm{diag}(\bm A)\ \mathrm{diag}(\bm l_0^{-1})\ (\bm l - \bm l_0)\,,
\end{equation}
where $\bm l_0$ is the rest length vector before the application of prestress. Thus, by adjusting $\bm l_0$ and $\bm l$, we can achieve different levels of prestress in the structure. Finally, the force density vector $\bm q \in\mathbb{R}^{M}$ reads
\begin{equation}
    \bm q = [q_{1} \ q_{2} \ \dots \  q_{M}]^\mathrm{T} 
          = \mathrm{diag}(\bm l^{-1})\ \bm t 
          = \mathrm{diag}(\bm E)\ \mathrm{diag}(\bm A)\ ( \bm l_0^{-1}- \bm l^{-1})\,,
    \label{eqn:q}
\end{equation}
reflecting geometric non-linearity through the current length vector $\bm{l}$.

Due to the nature of the problems under study, we ignore gravitational effects on the elements, focusing on the static response. Hence, the tensegrity statics can be formulated as
\begin{equation}
    \bm K \bm n = \bm f_\mathrm{ex}\,,
    \label{kn=f}
\end{equation}
where $\bm K$ is the stiffness matrix and $\bm f_\mathrm{ex}$ is the external force vector. The stiffness matrix $\bm K \in \mathbb{R}^{3N \bigtimes 3N }$ can be constructed from the force density vector and the connectivity matrix as
\begin{equation}
    \bm K = (\bm C^\mathrm{T}\ \mathrm{diag}(\bm q) \  \bm C) \otimes \mathbb{1}\,,
\end{equation}
where $\mathbb{1}$ is the identity matrix.
Accordingly, the incremental equilibrium equation reads
\begin{equation}
    \mathrm{d}(\bm K \bm n) = \mathrm{d}\bm f_\mathrm{ex} \,.
    \label{eqn:linearized}
\end{equation}
Since $\bm K$ is dependent on nodal coordinates $\bm n$, the left hand side of Eqn.~\ref{eqn:linearized} is expanded using the chain rule,
\begin{equation}
    \left [ \frac{\partial (\bm K \bm n)}{\partial \bm n} \right ] \mathrm{d}\bm n 
    = \left ( \bm K + \left [ \frac{\partial \bm q}{\partial \bm n} \frac{\partial (\bm K \bm n )}{\partial \bm q} \right ] \right ) \mathrm{d} \bm n \,.
    \label{eqn: del kn}
\end{equation}

From Eqn.~\ref{eqn:q}, we can show that the partial derivative of $\bm q$ with respect to $\bm n$ is given by
\begin{equation}
    \frac{\partial \bm q}{\partial \bm n}
    = \bm A_\mathrm{eq}\, \mathrm{diag}(\bm l)^{-3}\, \mathrm{diag}(\bm A) \, \mathrm{diag}(\bm E) \,,
\end{equation}
where
\begin{equation}
    \bm A_{\mathrm{eq}} = (\bm C^\mathrm{T} \otimes \mathbb{1}) \, \mathrm{blkdiag}(\bm N \bm C^\mathrm{T})\,.
\end{equation}
We can further show that $\bm K \bm n$ can be written as $\bm A_\mathrm{eq} \bm q$. Accordingly, Eqn.~\ref{eqn:linearized} is simplified to
\begin{equation}
    (\bm K_\mathrm{G} + \bm K_\mathrm{E} ) \mathrm{d} {\bm{n}} = \mathrm{d}\bm f_\mathrm{ex} \,,
    \label{eqn:Kdn=df}
\end{equation}
where
\begin{equation}
    \bm K_\mathrm{G} = \bm K = (\bm C^\mathrm{T} \mathrm{diag}(\bm q) \, \bm C) \otimes \mathbb{1}\,,
    \label{eqn:geometric_stiffness}
\end{equation}
\begin{equation}
    \bm K_\mathrm{E} = \bm A_{\mathrm{eq}} \, \mathrm{diag}(\bm E)\,  \mathrm{diag}(\bm A) \, \mathrm{diag}(\bm l)^{-3} \bm A_{\mathrm{eq}}^\mathrm{T}\,. \\
\end{equation}
Here, $\bm K_\mathrm{G}$ represents the geometric stiffness matrix, governed by the structure's topology and member force density; $\bm K_\mathrm{E}$ is the material stiffness matrix, governed by the elements' axial stiffness; and $\bm A_{\mathrm{eq}} \in \mathbb{R}^{3N \bigtimes M }$ is the equilibrium matrix. For full details of the general tensegrity FEM framework and derivations, readers are referred to \cite{MA2022}.

\subsection{Designing self-stress}
\label{sec:self-stress}
We can show that the equilibrium equation can be written as 
\begin{equation}
    \bm A_{\mathrm{eq}} \bm q = \bm f_\mathrm{ex}\,,
    \label{Aq=f}
\end{equation}
where Eqn.~\ref{Aq=f} is a linear form with respect to force density $\bm q$. With equilibrium written in this form, one can design initial self-stress as follows. We first apply singular value decomposition (SVD) to $\bm A_{\mathrm{eq}}$:
\begin{equation}
    \bm A_{\mathrm{eq}} = \bm U \bm \Sigma \bm V^\mathrm{T}\,.
    \label{svd}
\end{equation}
From \cite{PELLEGRINO1993}, we know the independent states of self-stress can be calculated from the null space of $\bm A_{\mathrm{eq}}$. For $\bm A_{\mathrm{eq}} \in \mathbb{R}^{3N \bigtimes M }$ with rank $r$, the number of independent self-stress modes $s$ can be calculated as 
\begin{equation}
    s = M - r\,.
\end{equation}
When $s = 0$, the structure is statically determinate and has a unique solution for any given load $\bm f_\mathrm{ex}$. When $s > 0$, there is an $s$-dimensional infinity of solutions. In this case, $\bm V$ in Eqn.~\ref{svd} can be expressed as
\begin{equation}
    \bm V = \big[\bm v_1 \ \bm v_2 \ \dots \ \bm v_r \ \ \dashedvert \ \ \bm w_1 \  \bm w_2 \ \dots \ \bm w_s\big]\,,
\end{equation}
where $\bm w_i \ (i = 1,2,\dots,s)$ are $s$ independent self-stress modes that satisfy Eqn.~\ref{Aq=f}. Letting
\begin{equation}
    \bm V_1 = [\bm v_1 \ \bm v_2 \ \dots \ \bm v_r]\,,
\end{equation}
\begin{equation}
    \bm V_2 = [\bm w_1 \  \bm w_2 \ \dots\ \bm w_s]\,,
\end{equation}
we can calculate the self-stabilizing $\bm q_0$ given initial $\bm f_0$ as
\begin{equation}
    \bm q_0 = \bm A_{\mathrm{eq}}^+ \bm f_0 + \bm V_2 \bm z\,,
    \label{eqn:q0}
\end{equation}
where $\bm A_{\mathrm{eq}}^+$ is the pseudo-inverse of $\bm A$ and $\bm z \in\mathbb{R}^s$ is the self-stress coefficient.

In Section \ref{sec:3dtes}, we will introduce a type of tensegrity structure based on truncated octahedrons. The given structure has only one self-stress mode, i.e. $s = 1$. In this case, $\bm V_2 \in \mathbb{R}^{M\bigtimes 1}$ and $z$ is a scalar. 

Suppose we wish to specify the member forces in bars as $f_\mathrm{b}$. Then, the self-stress coefficient can be calculated as \cite{MA2019a}
\begin{equation}
    z = \frac{\frac{f_\mathrm{b}}{l_\mathrm{b}}\mathbb 1_{M_{\mathrm{b}}} - \bm I_\mathrm{b} \bm A_{\mathrm{eq}}^+ \bm f_0}{\bm I_\mathrm{b} \bm V_2}\,,
    \label{eqn:z}
\end{equation}
where $l_\mathrm{b}$ is the bar length, $\mathbb 1_{M_{\mathrm{b}}} \in \mathbb{R}^{M_{\mathrm{b}}\bigtimes 1}$ is a vector of 1 with a size of the number of bar elements and $\bm I_\mathrm{b} \in \mathbb{R}^{M_{\mathrm{b}}\bigtimes M}$ is a matrix to select bar elements,
\begin{equation}
    I_{kl} = \left\{ \begin{array}{cl}
                    1 &  \ \mathrm{if} \ e^l \ \mathrm{is\ a\ bar\ element}\,, \\
                    0 &  \ \mathrm{otherwise} \,.
                    \end{array} \right.
\end{equation}
Accordingly, the self-stabilizing member forces can be calculated by plugging $z$ into Eqn.~\ref{eqn:q0}:
\begin{equation}
    \bm t_0 = \bm q_0 \, \mathrm{diag}(\bm l)\,.
    \label{eqn:t_0}
\end{equation}

\subsection{3D tessellation}
\label{sec:3dtes}
Tensegrity structures remain in self-equilibrium, which requires special arrangements of the elements. There is a vast literature on the form-finding of eligible self-stabilizing tensegrity configurations \cite{Tibert2003,Li2010_HJGao,LI2010_MC}. However, constructing 3-dimensional tensegrity lattices has been difficult due to the low symmetry in common elementary tensegrity cells \cite{RIMOLI2017}. For our purpose of modeling 3D cell mechanics, we adopt the 3-dimensional tensegrity lattices from truncated octahedron elementary cells \cite{RIMOLI2017,Li2010_HJGao} due to their ability to achieve space-tiling translational symmetry.

\begin{figure}[t!]
      \centering
      {\includegraphics[clip, trim = 0cm 0cm 8cm 16cm, width=0.6\textwidth]{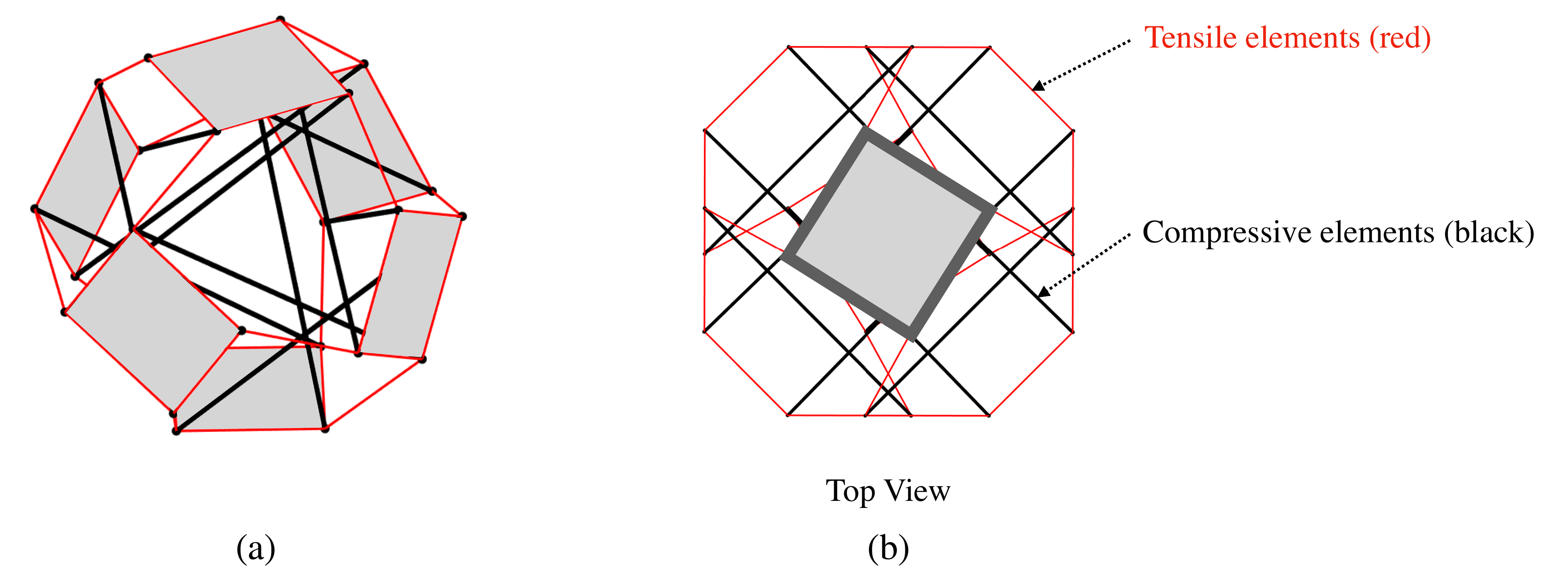} }
     \caption{(a) A tensegrity structure based on truncated octahedron elementary cell. Six side faces are filled with color for better viewing clarity. Side faces are twisted, as highlighted in darker contours in the (b) top view.}
 \label{fig:tenseg_elementary}
\end{figure}

\begin{figure}[t!]
      \centering
      {\includegraphics[clip, trim = 0cm 2cm 0cm 2cm, width=0.9\textwidth]{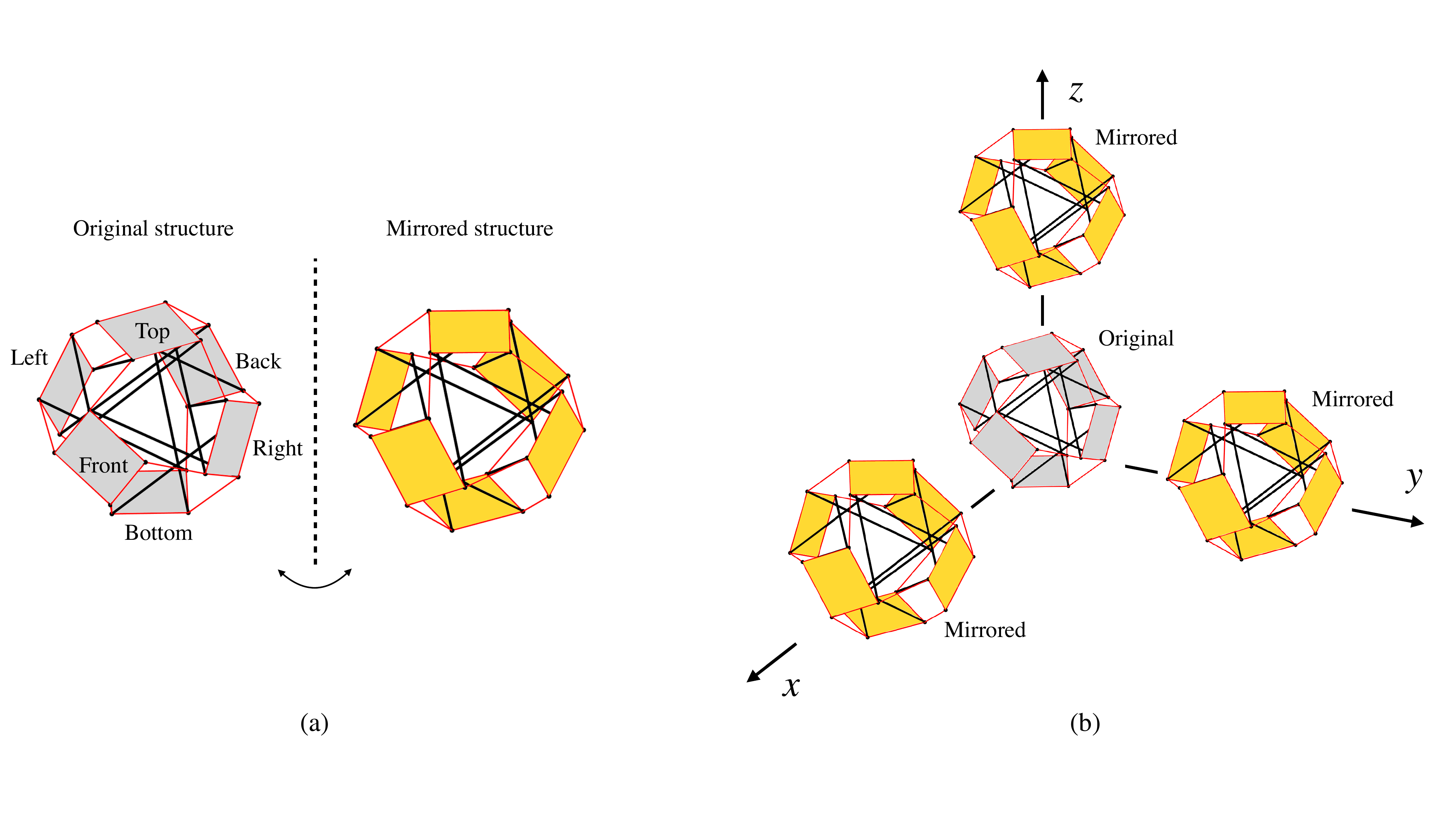} }
     \caption{Schematic showing how to connect neighboring cells.}
 \label{fig:tenseg_assembly}
\end{figure}

\begin{figure}[t!]
      \centering
      {\includegraphics[clip, trim = 2cm 4cm 8cm 8cm, width=0.6\textwidth]{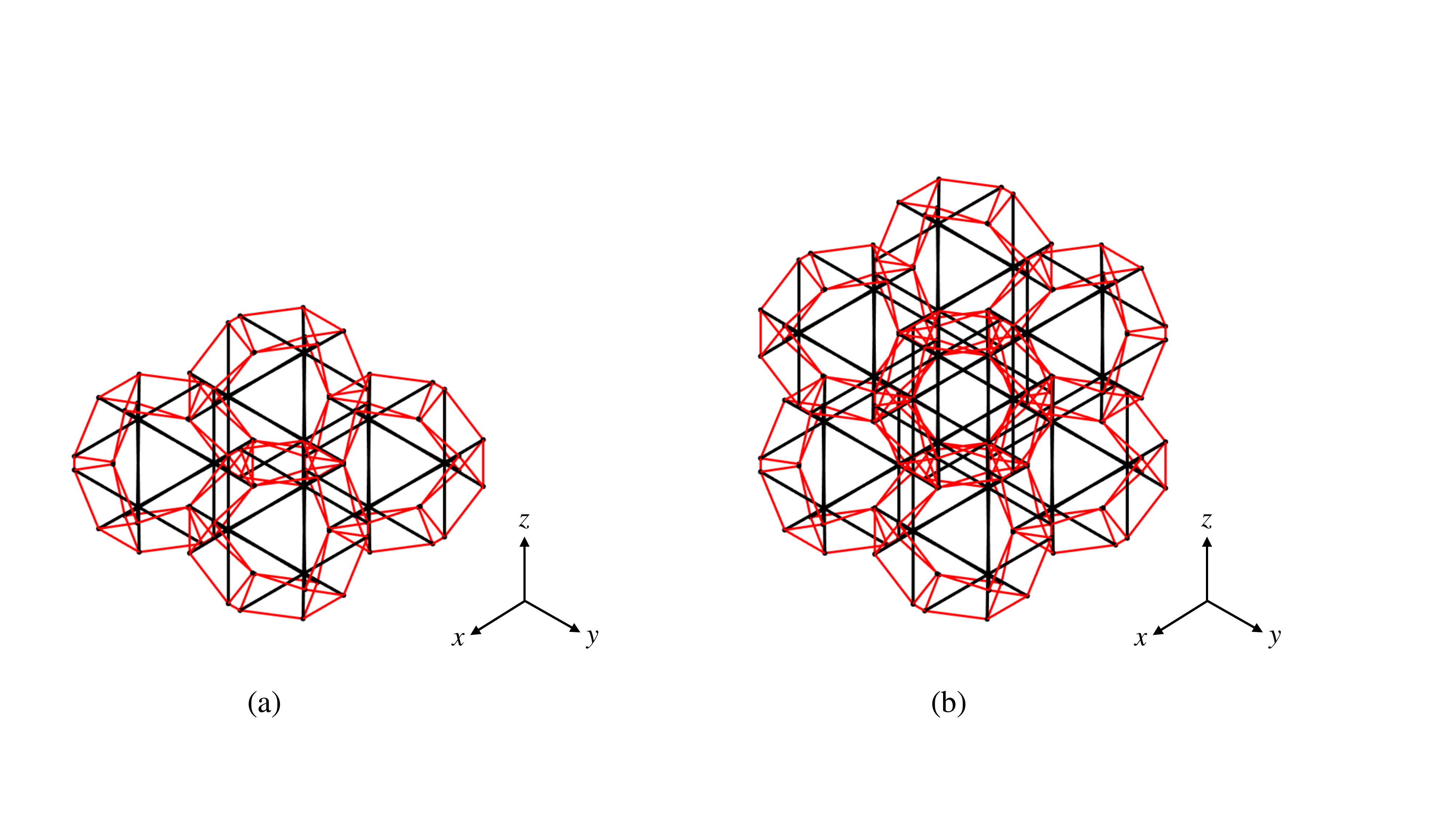} }
     \caption{(a) 2D and (b) 3D translational building blocks. }
 \label{fig:tenseg_2d3d}
\end{figure}

Each elementary unit is constructed from 12 compressive elements (bars, in black) and 36 tensile elements (tendons, in red), shown in Fig.~\ref{fig:tenseg_elementary}. The unit as it is right now is not 3D-translational or even 2D-translational due to the presence of twisted truncated surfaces (outlined in Fig.~\ref{fig:tenseg_elementary}b). For the structure to be translational in $\mathbb{R}^2$ and $\mathbb{R}^3$, the unit cell needs to be reflected with respect to any of the side surfaces for the twist to be aligned (Fig.~\ref{fig:tenseg_assembly}a). Any elementary unit cell must be paired up with reflected cells to make a connection, and no single pair of neighboring cells is identical (Fig.~\ref{fig:tenseg_assembly}b). By alternating reflected cells in $x$, $y$, and/or $z$ directions, the 4-cell system forms a building block that is translational in 2D (Fig.~\ref{fig:tenseg_2d3d}a), and an 8-cell system forms a building block that is translational in 3D (Fig.~\ref{fig:tenseg_2d3d}b) \cite{RIMOLI2017}. Note that the 2D translational blocks are still inherently 3D, as they possess a finite thickness.

\subsection{Extracting stresses and strains}
We are primarily concerned with stress and strain information representative of the global tensegrity structures.
Given a tensegrity assembly of volume $V$ with a total number of elements $M$, the average stress tensor can be calculated using the Christoffersen relation \cite{Christoffersen1981},
\begin{equation}
    \bar {\bm \sigma} = \frac{1}{V}\sum_{e=1}^{M} \mathrm{sym}\,(\bm t_e \otimes \bm l_e) \,,
\end{equation} 
where $\bm t_e$ is the internal force of element $e$ and $\bm l_e$ is the corresponding length vector.

Furthermore, the average strain can be calculated as \cite{Hohe2001}
\begin{equation}
    \bar{\bm \epsilon} = \frac{1}{V} \sum_{i \in J} \mathrm{sym}\,(\bm u_i  \otimes \hat{\bm n}_i) \,,
\end{equation}
where $J$ is the set of joint indices (node indices) on the boundary of the RVE, $\bm u = [u_x \ u_y \ u_z]^\mathrm{T}$ is the displacement of a node, and $\bm {\hat n}$ is the outward surface normal on the nodes contained in $J$.

\section{Modeling cell mechanics}
\label{sec:cell}
In the cell CSK, the tensile actin filaments and compressive microtubules are reminiscent of tensegrity structures \cite{Ingber2014, Ingber2023}. In this section, we simulate and capture the mechanics of a single cell, a monolayer, and a multicellular spheroid using the proposed 3D tensegrity structure.

\subsection{Single cell}
\label{sec:single_cell}
Harris et al.~\cite{Harris2011} experimentally measured the force-displacement response of epithelial cells using combined AFM–confocal microscopy, which provides a good benchmark to calibrate the parameters used in our tensegrity model. We start by constructing a single elementary unit tensegrity cell, as depicted in Fig.~\ref{fig:tenseg_elementary}. The height of this elementary unit is set to match the height of a single epithelial cell used in the aforementioned experiment. The tendons are used as simplified representations of the actin cortex and stress fibers, and bars as representations of the microtubules and stiffened cross-linked actin bundles \cite{SUN2023_feng}. The bottom nodes are fixed to mimic the experimental environment. Due to the large variability in reported parameters for CSK components, such as the modulus and diameter of filaments, we iteratively calibrate these values, staying within the ranges reported in \cite{Ingber2014}, until the model’s stress–strain response aligns with the range of experimental data observed in the benchmark study. Final model parameters are listed in Table~\ref{tab:single_cell}. 

\begin{figure}[t!]
      \centering
      {\includegraphics[clip, trim = 0cm 0cm 26cm 6cm, width=0.5\textwidth]{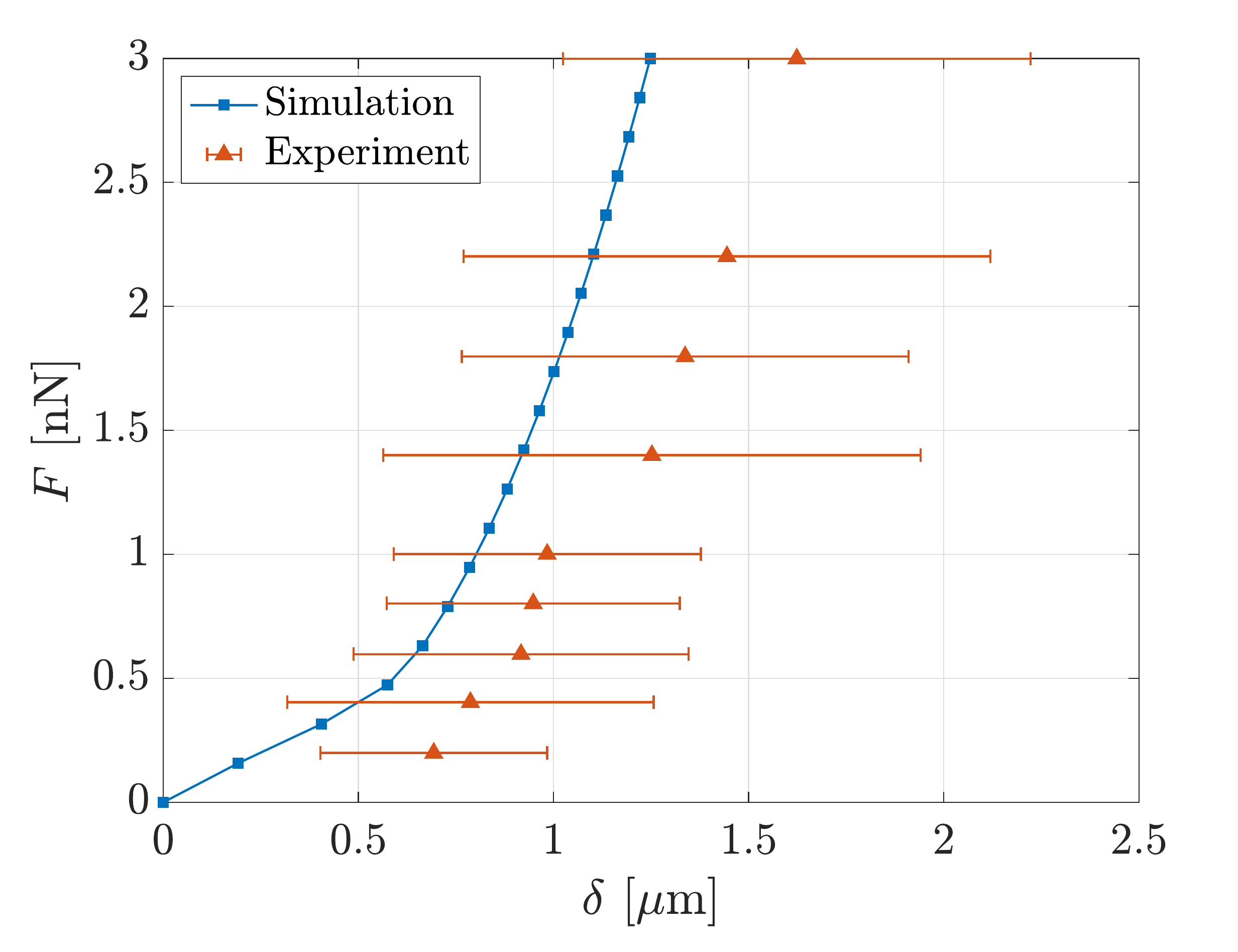} }
     \caption{Force-displacement plot of a simulated single-cell indentation test. Experimental data is adapted from \cite{Harris2011}.}
 \label{fig:single_cell_plot}
\end{figure}

\begin{figure}[t!]
      \centering
      {\includegraphics[clip, trim = 2cm 0cm 2cm 0cm, width=0.6\textwidth]{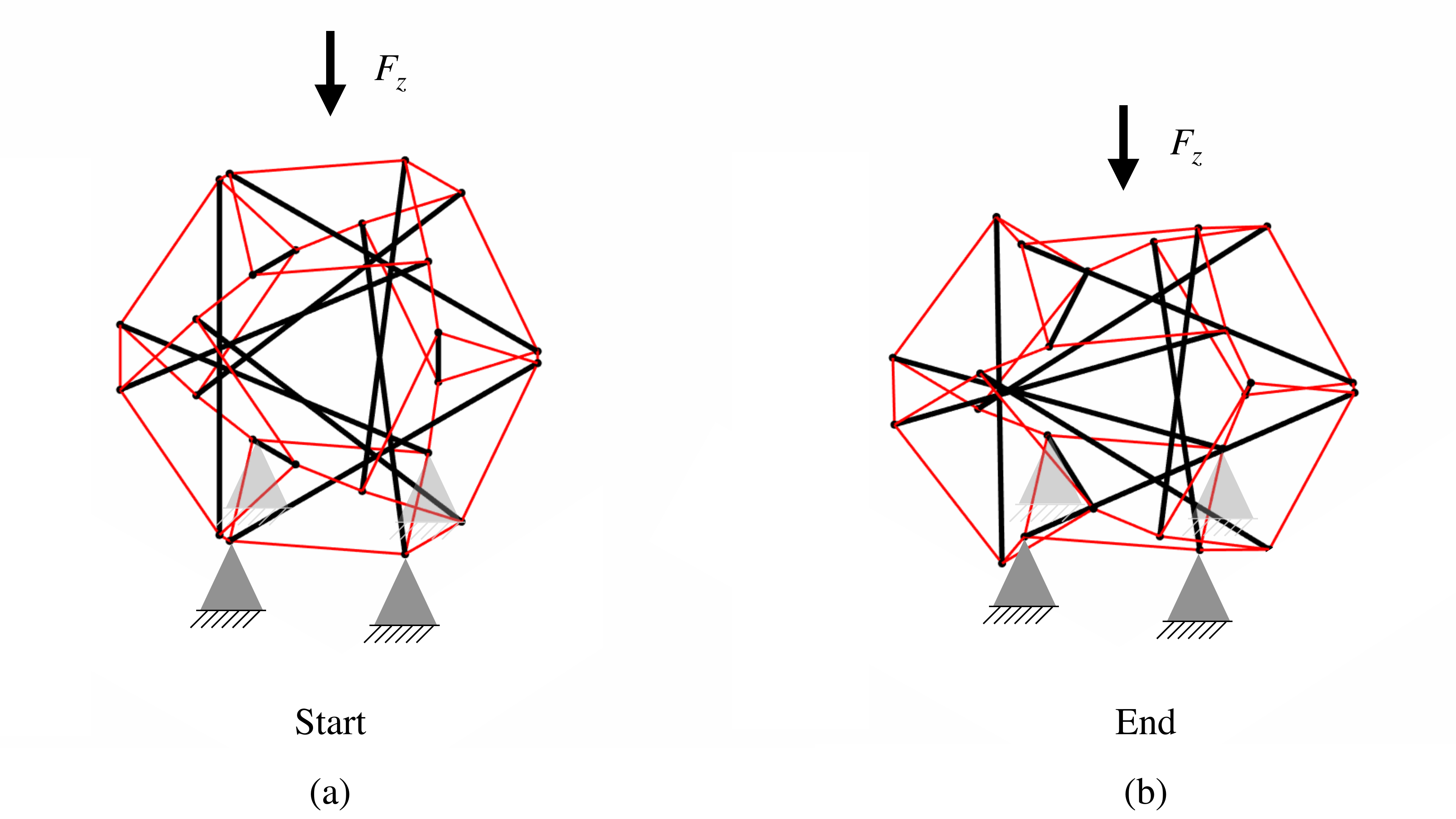} }
     \caption{Single cell (a) before and (b) after the indentation test.}
 \label{fig:single_cell_ani}
\end{figure}

An indentation force of up to 3 nN, same as the experiment, is applied to the top nodes, as shown in Fig.~\ref{fig:single_cell_ani}. From the resulting force-displacement plot shown in Fig.~\ref{fig:single_cell_plot}, we see that a single elementary unit using the proposed tensegrity structure captures the force-displacement profile of an epithelial cell in an indentation test. As the indentation force increases, the rate of change in indentation displacement slows down, demonstrating nonlinearity in the cell response. This benchmark also serves as a calibration for the model parameters, which will be consistently used in subsequent multicellular tests.

\begin{table}[h]
        \small
        \centering
        \vspace{1em}
        \caption{Parameters used in indentation simulation}
        \begin{tabular}{lcll}
        Parameter                         & Value     & Units & Reference \\ \hline
        Axial stiffness in bars $E_\mathrm{b}$     &   1.2  &  GPa   & \cite{Gittes1993} \\     
        Axial stiffness in tendons $E_\mathrm{s}$     & 10   &  MPa     & Assumed from \cite{Ingber2014} \\
        Diameter of bars  &  0.1        & $\mu$m       & Assumed from \cite{Ingber2014} \\
        Diameter of tendons &  1    & $\mu$m   & Assumed from \cite{Ingber2014} \\ 
        Applied force $F$                 &  3        &   nN  &  \cite{Harris2011} \\
        Cell height  $h$            & 10        & $\mu m$  & \cite{Harris2011} \\ \hline   
        \end{tabular}
        \label{tab:single_cell}
\end{table}


\subsection{Monolayer}
\label{sec:monolayer}
Harris et al.~\cite{Harris2012} reported the mechanical behavior of a freely suspended epithelial cell monolayer under uniaxial stretch and observed that the modulus of a monolayer is about two orders of magnitude larger than that of an isolated cell reported in \cite{Harris2011}. To capture this distinction, we construct a tensegrity monolayer made of 60$\times$30$\times$1 2D translational building blocks (Fig.~\ref{fig:tenseg_2d3d}a) with the same model parameters as Section \ref{sec:single_cell} (Table~\ref{tab:single_cell}). This domain size is determined to be sufficiently large to produce a representative mechanical response under uniaxial loading, based on convergence tests where further increasing the monolayer dimensions results in negligible changes in the stress–strain behavior. To replicate the experimental condition, we fix all the nodes on the bottom side of the monolayer and prescribe a 25\% strain on the top side. 

\begin{figure}[t!]
      \centering
      {\includegraphics[clip, trim = 0cm 0cm 26cm 6cm, width=0.5\textwidth]{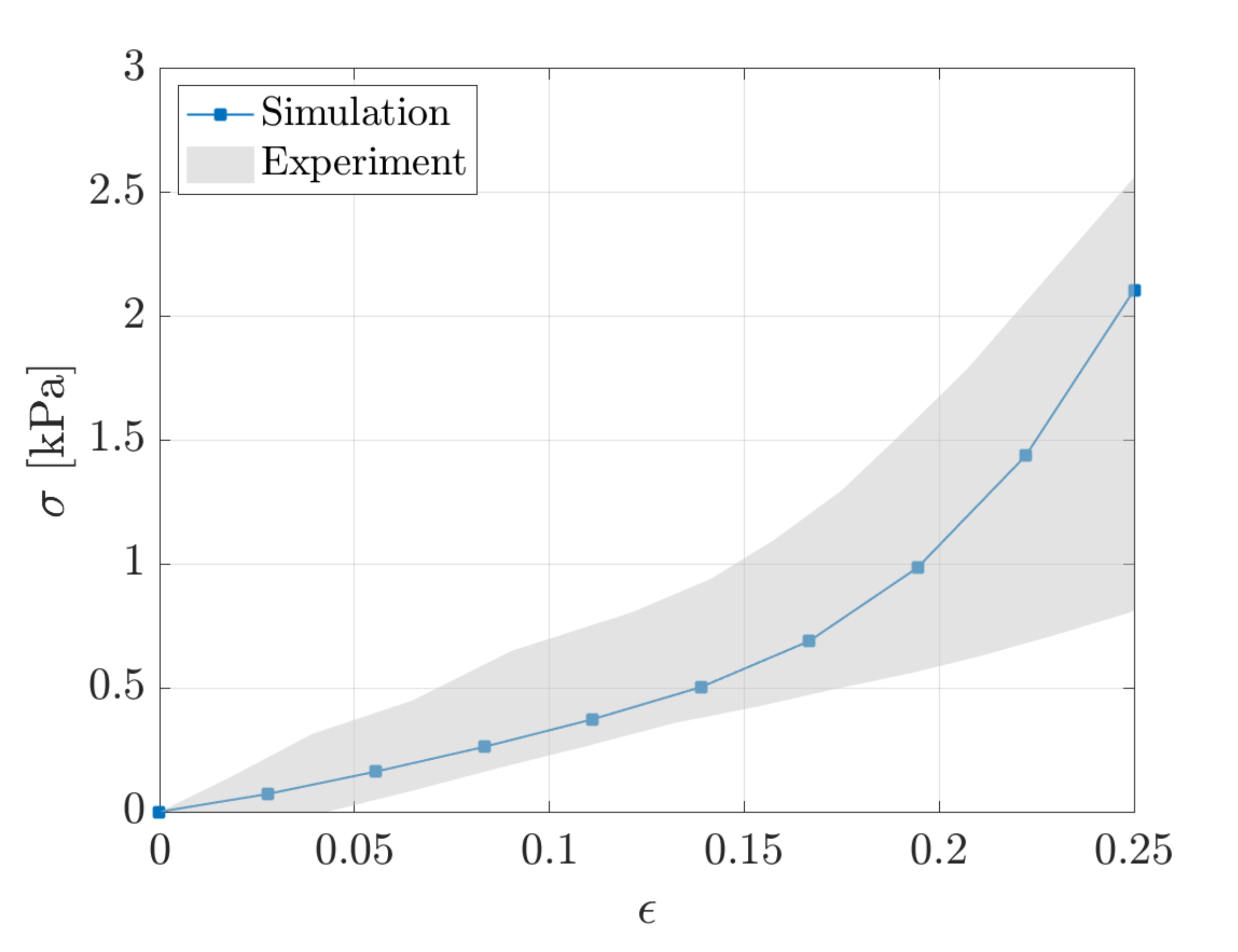} }
     \caption{Stress-strain plot of the simulated monolayer uniaxial stretch up to 25\% strain. Experimental data is adapted from \cite{Harris2012}. The envelope shows the maximum and minimum stress values taken from the experiment at a given strain.}
 \label{fig:mono_cell_plot}
\end{figure}

The resulting stress-strain plot is shown in Fig.~\ref{fig:mono_cell_plot}. We see that the simulated response of 25\% stretch falls within the range of variability of the reported experimental data. Up to 25\% strain, we observe a nonlinear region where the rate of change of stress increases slowly. Since we assume the bars and tendons are linear elastic, this onset of nonlinearity below 25\% can only come from the geometric stiffness of the structure. Upon loading, members of the tensegrity structure geometrically realign in the loading direction, therefore kinematically stiffening the structural response \cite{Ingber2014}. 

Beyond 25\%, Harris et al.~\cite{Harris2012} reported that stress increases linearly in the 25\% to 50\% strain range. Mechanical failure of the monolayer is reported to occur at a strain value over 70\%. In our simulation, the degree of stiffness at higher strains diverges from these observations: the simulated response above 25\% becomes too stiff and falls out of the experimental envelope. We suspect the reason behind this discrepancy is related to the model's simplified geometry and constitutive assumptions, making it suboptimal to capture the complex cellular-level mechanics occurring at large strains, such as intercellular adhesion rupture. These differences highlight the limitations of the current model in capturing the full mechanical complexity of real cells, particularly under large tensile strains. While more advanced constitutive relations, such as deformation-dependent material laws, may better capture the nonlinear nature of the CSK, they typically involve additional numerical parameters that require tuning and often lack experimental grounding. Therefore, we adopt a linear stress-strain relation (with inherent geometric nonlinearity) to maintain model tractability and interpretability. Nonetheless, the tensegrity framework captures key qualitative features, such as strain-stiffening behavior, and provides an interpretable, structurally grounded framework for representing cellular~mechanics.

\begin{figure}[t!]
      \centering
      {\includegraphics[clip, trim = 2cm 0cm 0cm 2cm, width=0.7\textwidth]{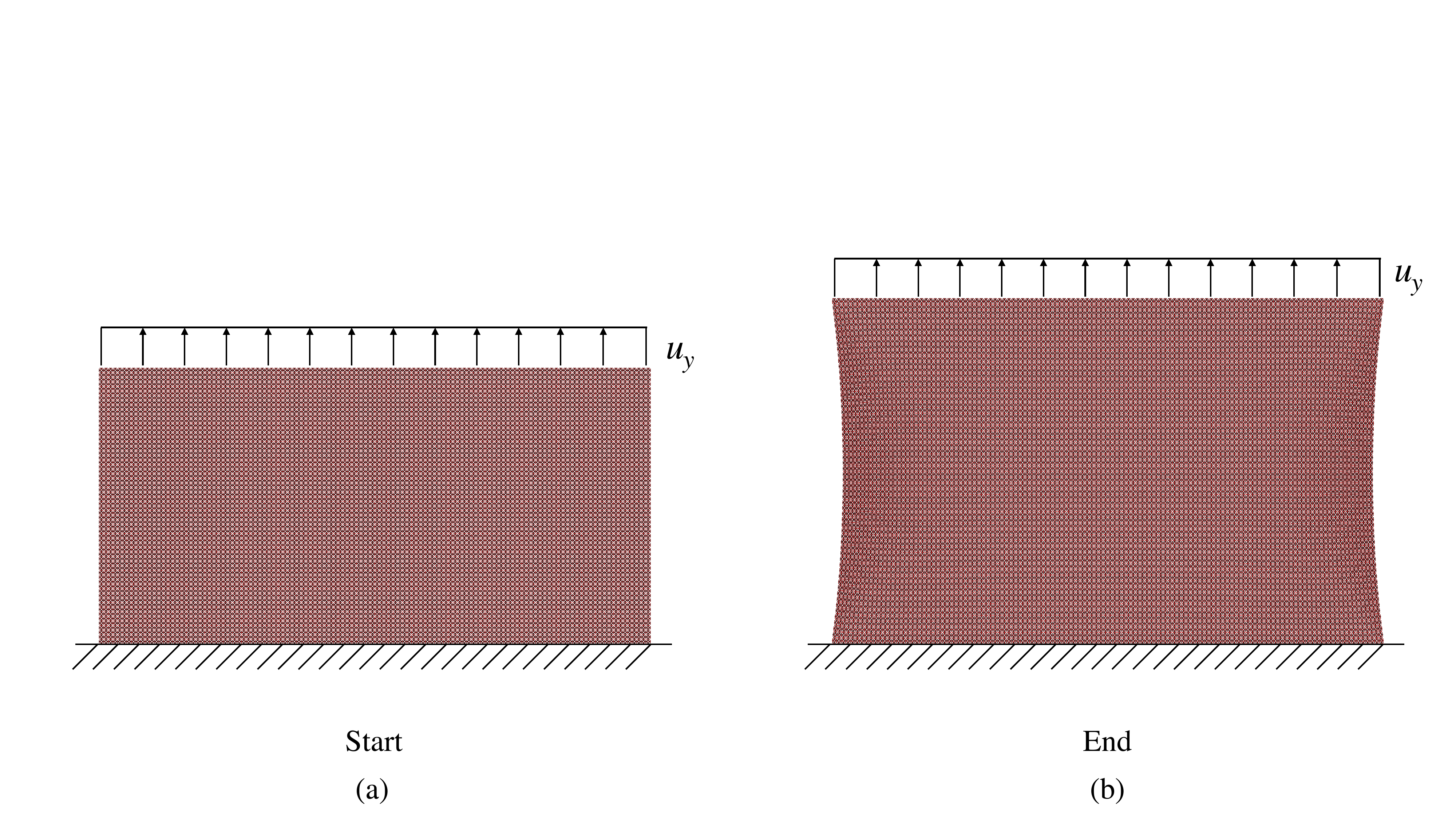} }
     \caption{Monolayer (a) before and (b) after the tensegrity monolayer uniaxial stretch.}
 \label{fig:mono_cell_ani}
\end{figure}

\begin{figure}[t!]
      \centering
      {\includegraphics[clip, trim = 0cm 0cm 0cm 10cm, width=1\textwidth]{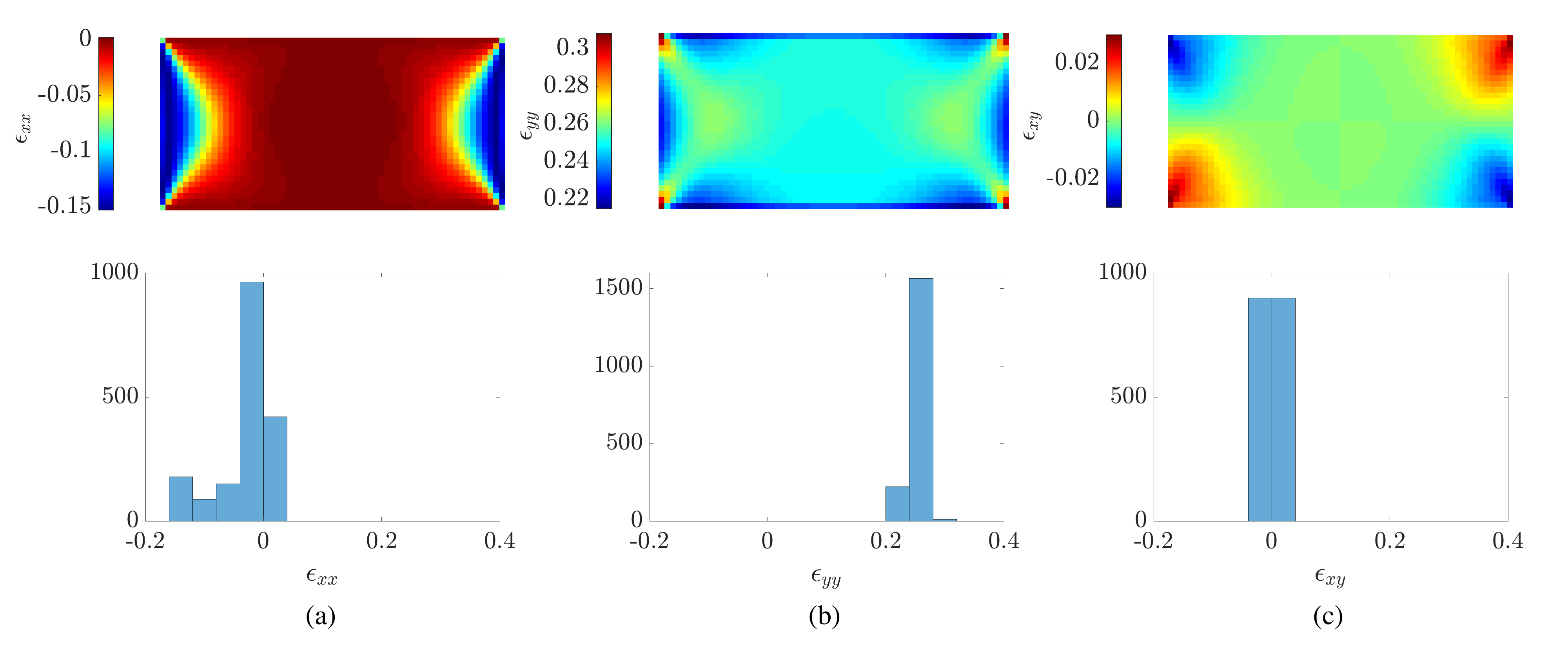} }
     \caption{Visualization of the strain fields and histogram plots of $\epsilon_{xx}$, $\epsilon_{yy}$, and $\epsilon_{xy}$ at 25\% strain of the monolayer stretch test.}
 \label{fig:mono_cell_hist}
\end{figure}

Using our model, we can easily extract the stress and strain field information from the tensegrity monolayer. In Fig.~\ref{fig:mono_cell_hist}, we plot the $\epsilon_{xx}$, $\epsilon_{yy}$, and $\epsilon_{xy}$ strain components of the tensegrity monolayer at the end of the 25\% stretch. From the histograms, we see that the normal strain $\epsilon_{yy}$ is centered around 0.25, showing a quasi-uniform strain field in the $y$ direction. On the other hand, the shear strain $\epsilon_{xy}$ is centered around 0. Most of the $\epsilon_{xx}$ strain is centered around 0, with a few values around $-0.1$. From the visualization (Fig.~\ref{fig:mono_cell_ani}), we see that the tensegrity cells near the left and right edge of the monolayer curve inwards, which is often the case for the geometry of this shape. These results have also been reported in \cite{Harris2012}, showing an overall good agreement between our simulation and the published experimental data.

So far, we have demonstrated that our proposed tensegrity model can capture the nonlinear behavior of a single cell in an indentation test and of a monolayer in a uniaxial tension test at low (but not necessarily infinitesimal) strain. In the next subsection, we will show that our model can also be constructed to capture 3D cell structures, a capability not previously documented in the literature for applications in cell mechanics simulations. We will provide an example illustrating how this 3D structure can be used to capture the mechanics of a multicellular spheroid.

\subsection{Multicellular spheroid}
Studies show that tumor growth under constrained conditions decreases cell proliferation and induces apoptosis in a non-uniform manner \cite{Helmlinger1997,Delarue2014,Cheng2009}. Spherical aggregates of cells mimicking a tumor, or multicellular spheroids (MCSs), are reported to exhibit a two- to six-fold increase in cell density near their geometric center \cite{Valencia2015}. This heightened crowding in the subcellular environment can result in a decrease in cellular rearrangement liberty, which in turn correlates to hampered proliferation \cite{Dolega2017}.

To understand such cellular phenotypic heterogeneity, Dolega et al.~\cite{Dolega2017} experimentally studied the stress distribution inside MCSs of malignant murine colon cancer cells under osmotic pressure, mimicking a growth-constrained environment. Their study shows nontrivial stress patterns within the MCS, with a pressure rise towards the core, thus suggesting a direct link to the lack of proliferation near the spheroid's center. 

To capture such a heterogeneous stress distribution, we construct a 3D tensegrity model with non-uniform prestress to replicate the results observed in \cite{Dolega2017}, and to relate to the non-uniform cell proliferation reported in \cite{Montel2012}. In our 3D tensegrity model, we prescribe cellular prestress to mimic the crowding-induced stiffening effect \cite{Zhou2009} near the MCS core. 

\begin{table}[h]
        \small
        \centering
        \vspace{1em}
        \caption{MCS model construction details}
        \begin{tabular}{lc}
        Parameter                         & Value     \\ \hline
        Number of 3D translational units  &   552  \\ 
        Number of elementary units  &   4416  \\ 
        Number of bars $M_\mathrm{b}$  &   52992  \\ 
        Number of tendons $M_\mathrm{s}$  &   109824  \\ 
        Number of nodes $N$  &   56832  \\ \hline
        \end{tabular}
        \label{tab:spheroid}
\end{table}

\begin{figure}[t!]
      \centering
      {\includegraphics[clip, trim = 0cm 0cm 10cm 10cm, width=0.6\textwidth]{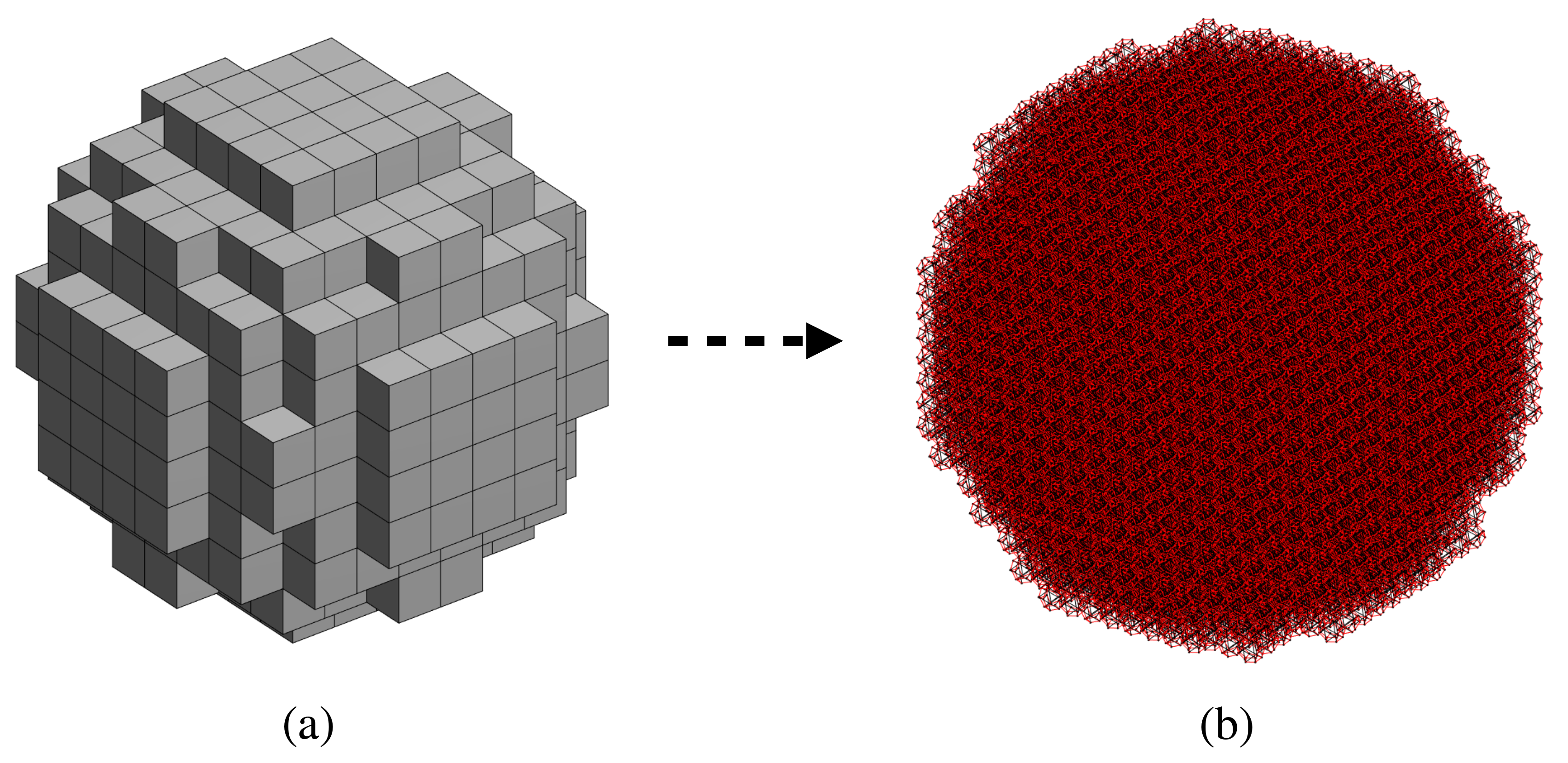} }
     \caption{A spheroid constructed using 3D tensegrity building blocks. The gray voxelated sphere shown in (a) illustrates how the spheroid is initially constructed. Each individual block in this voxelated sphere is then replaced with a 3D tensegrity building block to create the final structure shown in (b).}
 \label{fig:spheroid}
\end{figure}

\begin{figure}[t!]
      \centering
      {\includegraphics[clip, trim = 0cm 0cm 20cm 0cm, width=0.5\textwidth]{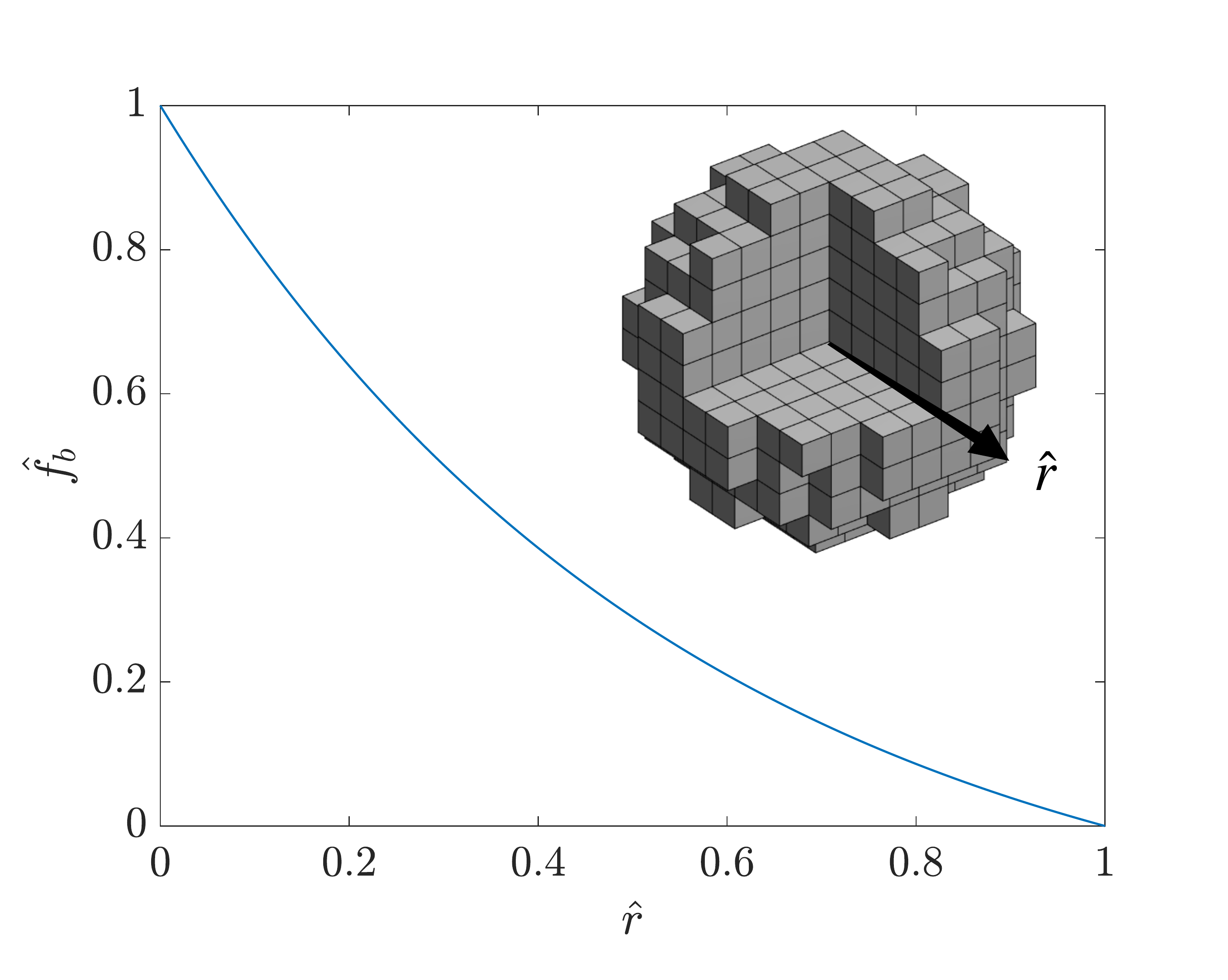} }
     \caption{Normalized initial bar force vs. normalized radial position. Bar forces are normalized with respect to the maximal bar force, which is at the center of the MCS. Radial positions are normalized by the radius of the MCS. We assume an exponential relation of the initial bar force, based on the exponential decay of distribution of cells within spheroids \cite{Valencia2015}. }
 \label{fig:prestress}
\end{figure}

We initiate a voxelated sphere-like aggregate as shown in Fig.~\ref{fig:spheroid} with structural details reported in Table~\ref{tab:spheroid}. Each cube from Fig.~\ref{fig:spheroid}a represents a 3D translational building block in Fig.~\ref{fig:tenseg_2d3d}b. To simulate more cellular stiffening towards the core, we assume increasing compressive forces in bar elements for each building block as a function of the radial position of the block within the MCS (Fig.~\ref{fig:prestress}). The balancing forces in the rest of the elements are calculated via procedures explained in Section~\ref{sec:self-stress}. This non-uniform assignment of initial forcing in each building block consequently changes the stiffness across the MCS according to Eqns.~\ref{eqn:Kdn=df} and~\ref{eqn:geometric_stiffness}.

We then prescribe an isotropic stress to the MCS and plot the normalized radial stress profile $\hat \sigma_{rr}$ with respect to the normalized radial position $\hat r$. Note that although the internal prestress is assigned according to a spatially varying profile, each individual
tensegrity cell is initially in a self-equilibrated state, meaning that all internal forces are balanced without
requiring any external loads or interactions with neighboring cells. Once external compression is applied, the system departs from its initial self-equilibrated state.
As shown in Eqn.~\ref{eqn:Kdn=df}, external loading alters the force densities $\bm q$, which in turn modify the geometric stiffness matrix $\bm K_\mathrm{G}$. This coupling causes nonlinear and spatially dependent mechanical responses that emerge from both the structural geometry and prestress distribution. Therefore, although $f_\mathrm{b}$ is prescribed, the resulting mechanical behavior under compression is emergent and nontrivial.

\begin{figure}[t!]
      \centering
      {\includegraphics[clip, trim = 0cm 0cm 0cm 10cm, width=0.8\textwidth]{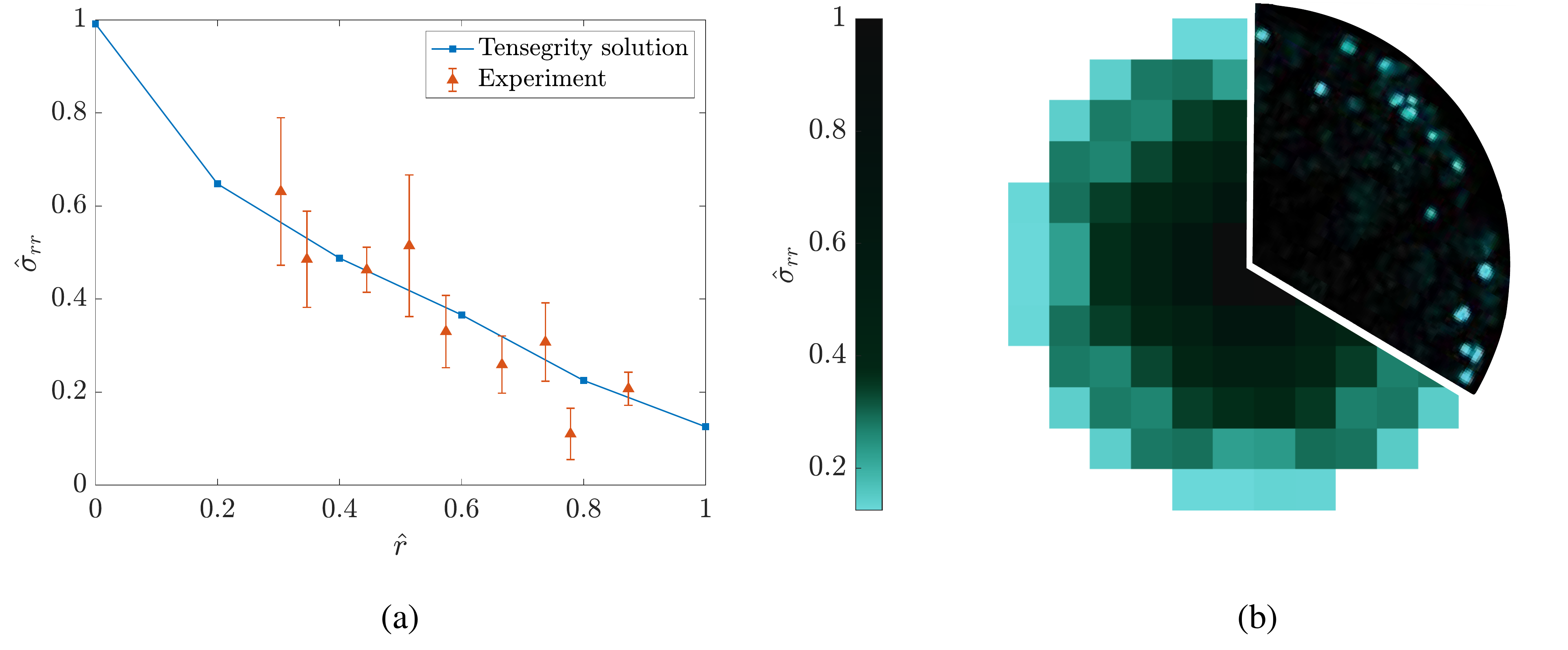} }
     \caption{(a) Normalized radial stress vs.~normalized radial position. $\sigma_{rr}$ is normalized with respect to the maximum value. Experimental data is adapted from \cite{Dolega2017} and scaled to match with the simulated data. (b) Visualization of the normalized radial stress at the middle slice of the MCS. The top right image shows cellular proliferation along the radius in a spheroid grown under constraint, with Ki-67 immunostained \cite{Montel2012}.} 
 \label{fig:proliferation}
\end{figure}

As seen in Fig.~\ref{fig:proliferation}a, upon an assumed initial bar force specification $f_\mathrm{b}$ and corresponding member forces calculated using Eqns.~\ref{eqn:t_0} and~\ref{eqn:z}, the resulting rate of change of $\hat \sigma_{rr}$ is in good alignment with the experimental result. In Fig.~\ref{fig:proliferation}b, we visualize the normalized radial stress of the middle slice of the MCS and overlay it with an experimental image from \cite{Montel2012}, where Ki-67 (a proliferation marker) is immunostained. We observe that the regions of high radial stress near the center of the numerical MCS overlap well with the low proliferation regions within the experimentally tested cell, visually highlighting the clear correlation between stress and cell proliferation. 

Rather than serving as an independent validation in the traditional experimental sense, this result is intended to demonstrate that the tensegrity-based cell model can incorporate internal prestress and capture its mechanical implications. This result thus suggests that our model can be used to simulate the stress distribution within a 3D MCS, showcasing its potential as a tool for studying the correlative effects of mechanical behavior on cellular functionalities. 

\subsection{Discussion}
Overall, we observe good agreement between our simulated results and published experimental data using the proposed tensegrity model, particularly at low (but not infinitesimal) strain values. However, at high strain values, such as those exceeding 25\% in the uniaxial tension test, the simulated results fall out of the experimental envelope. There could be multiple factors contributing to this discrepancy. One factor is that our model could be oversimplified and based on limiting assumptions. For example, we assume the bars and tendons are linearly elastic, whereas, in reality, actin filaments and microtubules are reported to exhibit nonlinear material behaviors \cite{Ingber2014}. Additionally, the present model cannot simulate the complex cellular-level mechanics occurring at high strain values, such as detachment of focal adhesion, cell rupture, etc.

Since we use pin-pin connections between neighboring tensegrity cells, our model must be constructed in an orderly manner, which prevents anisotropic orientation in the initiation of the cells. Additionally, the shape of the elementary tensegrity unit needs to be carefully designed to ensure that its members reach equilibrium. This is a disadvantage shared by all tensegrity-based cell models. However, by relating the cell CSK to bars and tendons, we can individually assign different stiffnesses, cross-sectional areas, or even prestress levels to each tensegrity cell. This approach allows us to study the overall response of tissues while locally modulating the properties of the cell CSK.
		
\section{Multiscale data-driven computing with tensegrity}
\label{sec:DD}
\FloatBarrier

Living organisms are made of millions of cells \cite{Bianconi2013}. Even the monolayer used in \cite{Harris2012} was cultured using 25000 cells. If we were to represent each cell using any abstract discrete model, be it tensegrity or others, the computational cost would increase exponentially as we increase the scale of the problem. Direct numerical simulations at full resolution can potentially pose a significant computational burden, especially considering (large) nonlinear stiffness matrices, often requiring many iterations to converge. This limitation sets back the simplicity in simulating cell mechanics using abstract models such as the proposed tensegrity structures.

Aiming to address this problem, we adopt the data-driven (DD) computing paradigm proposed by Kirchdoerfer and Ortiz~\cite{KIRCHDOERFER2016}, which serves as a multiscale analysis tool~\cite{KARAPIPERIS2021} in the general context of solid mechanics. This paradigm bypasses the definition of a constitutive relation, for instance, relating stress to strain given empirical material properties. Given existing material datasets, the DD problem is formulated as a minimal distance search subject to the fundamental equilibrium and compatibility constraints. The material data can come from either simulation or experiments, rendering the DD framework free of constitutive relations, efficient in computation, and versatile in application. Since then, the DD framework has been extended to dynamics \cite{Kirchdoerfer2018}, fracture \cite{CARRARA2020}, finite deformation \cite{NGUYEN2018}, micromorphic continua \cite{ULLOA2024b}, etc. The multiscale approach of the DD framework has been presented in the analysis of history-dependent materials \cite{KARAPIPERIS2021} and breakage mechanics \cite{ULLOA2023}, showing good agreement with direct numerical simulations of physics-based models. DD computing has also been applied to calculate the mechanical response of the human brain from \textit{in situ} and \textit{in vivo} imaging data \cite{Salahshoor2024}.

In order to tackle the heavy computational burden of simulating large-scale cell mechanics using tensegrity structures, we perform a multiscale DD analysis on a monolayer with data extracted from lower-scale simulations of discrete tensegrity FEM models. Due to the limited capacity of the tensegrity model to accurately capture large deformations in cells, as discussed in Section~\ref{sec:monolayer}, the present section is restricted to small deformations. 

\subsection{Modeling framework}
We briefly summarize the infinitesimal deformation DD framework as follows. Given a body in $\mathbb R^\mathrm{d}$ discretized into $N$ nodes and $M$ material points, subject to external nodal forces $\bm f_i\in \mathbb R^\mathrm{d} \ (i = 1,2,\dots,N) $, and exhibiting nodal displacements $\bm u_i\in \mathbb R^\mathrm{d} \ (i = 1,2,\dots,N)$, the internal state is characterized by stress-strain pairs $(\bm \sigma_e,\bm \epsilon_e)$ at each material point $e, \  e = 1,2,\dots,M$. Each stress-strain pair is referred to as a local phase-space coordinate $\bm z_e\in Z_e$, where the global phase space reads $Z=Z_1\times Z_2\times\dots\times Z_M$.

Within the FEM setting, the internal state at any specific time $t_k$ is constrained by the following equilibrium and compatibility equations:
\begin{equation}
    \bm \epsilon_e^k = \bm B_e \bm u^k,\ e = 1,2,\dots,M\,,
    \label{eqn:compatibility}
\end{equation}
\begin{equation}
    \sum_{e=1}^M w_e \bm B_e^\mathrm{T} \bm \sigma_e^k = \bm f^k\,,
    \label{eqn:equilibrium}
\end{equation}
which define the constraint set $E_k$ containing the \emph{mechanical solution} at $t_k$:
\begin{equation}
    E_k = \{\bm z \in Z \ | \ \mathrm{(\ref{eqn:compatibility}) \ and\  (\ref{eqn:equilibrium})} \}\,.
\end{equation}

Instead of relating $\bm \sigma_e$ to $\bm \epsilon_e$ locally using a constitutive relation, DD framework introduces the global minimization problem between the constraint set $E$ and an existing material dataset $D \subset Z$ containing the \emph{material solution}. The distance metric can be written as 
\begin{equation}
    \| \bm z \| = \left ( \sum_{e = 1}^M w_e \| \bm z_e \|^2\right )^\frac{1}{2}\,,
    \label{eqn:z}
\end{equation}
with the local distance
\begin{equation}
    \| \bm z_e\| = \left (\frac{1}{2} \bm \epsilon_e : \mathbb C_e : \bm \epsilon_e + \frac{1}{2} \bm \sigma_e : \mathbb C_e^{-1} : \bm \sigma_e \right )^\frac{1}{2}\,.
    \label{eqn:ze}
\end{equation}
Here, $w_e$ is the weight associated with each material point, and $\mathbb C_e$ is a symmetric matrix serving as a purely numerical operator. The time-discrete DD problem can then be written as
\begin{equation}
    \inf_{\bm y \in D_k}  \inf_{\bm z \in E_k} \| \bm y - \bm z \|^2 =  \inf_{\bm z \in E_k}  \inf_{\bm y \in D_k} \| \bm y - \bm z \|^2\,.
\end{equation}
This problem can be solved using fixed-point iterations, i.e., by iteratively finding the closest point projection $(\bm \sigma_e,\bm \epsilon_e)$ onto the hyperplane that satisfies Eqns.~\ref{eqn:compatibility} and~\ref{eqn:equilibrium}, and then finding the closest material data point $(\bm \sigma_e^*,\bm \epsilon_e^*)$. The closest point projection can be resolved using a nodal set of Lagrange multipliers $\bm \eta$. Then, the FEM  equations of the DD problem take the form
\begin{equation}
    \left (\sum_{e=1}^M w_e\bm B_e^\mathrm{T} \mathbb C_e \bm B_e \right ) \bm u  = \sum_{e=1}^M w_e\bm B_e^\mathrm{T} \mathbb C_e \bm \epsilon_e^* \,,
\end{equation}
\begin{equation}
    \left (\sum_{e=1}^M w_e\bm B_e^\mathrm{T} \mathbb C_e \bm B_e \right ) \bm \eta  = \bm f - \sum_{e=1}^M w_e\bm B_e^\mathrm{T} \bm \sigma_e^* \,,
\end{equation}
\begin{equation}
    \bm \sigma_e = \bm \sigma_e^* + \mathbb C_e \sum_{i=1}^N \bm B_{ei} \bm \eta_i \,.
\end{equation}
Readers are referred to, e.g., \cite{KIRCHDOERFER2016,KARAPIPERIS2021} for further details.

\subsection{Monolayer tests}
\label{sec:data_sampling}
We proceed to use the DD framework explained above to capture the monolayer cell mechanics. We construct a tensegrity monolayer made of 60$\times$30$\times$1 3D translational building blocks (Fig.~\ref{fig:tenseg_2d3d}b). Compared to Section~\ref{sec:monolayer}, the thickness in the $z$-direction is increased through the use of 3D translational cells to provide better through-thickness representation for subsequent homogenization. In the DD solutions, we employ a 3D mesh with 60$\times$30$\times$1 elements, each with eight nodes and eight Gauss integration points. Noting the small-deformation setting, in this section, we consider monolayer tensile and simple shear tests up to 5\% strain.

\begin{figure}[t!]
      \centering
      {\includegraphics[clip, trim = 0cm 0cm 0cm 12cm, width=1\textwidth]{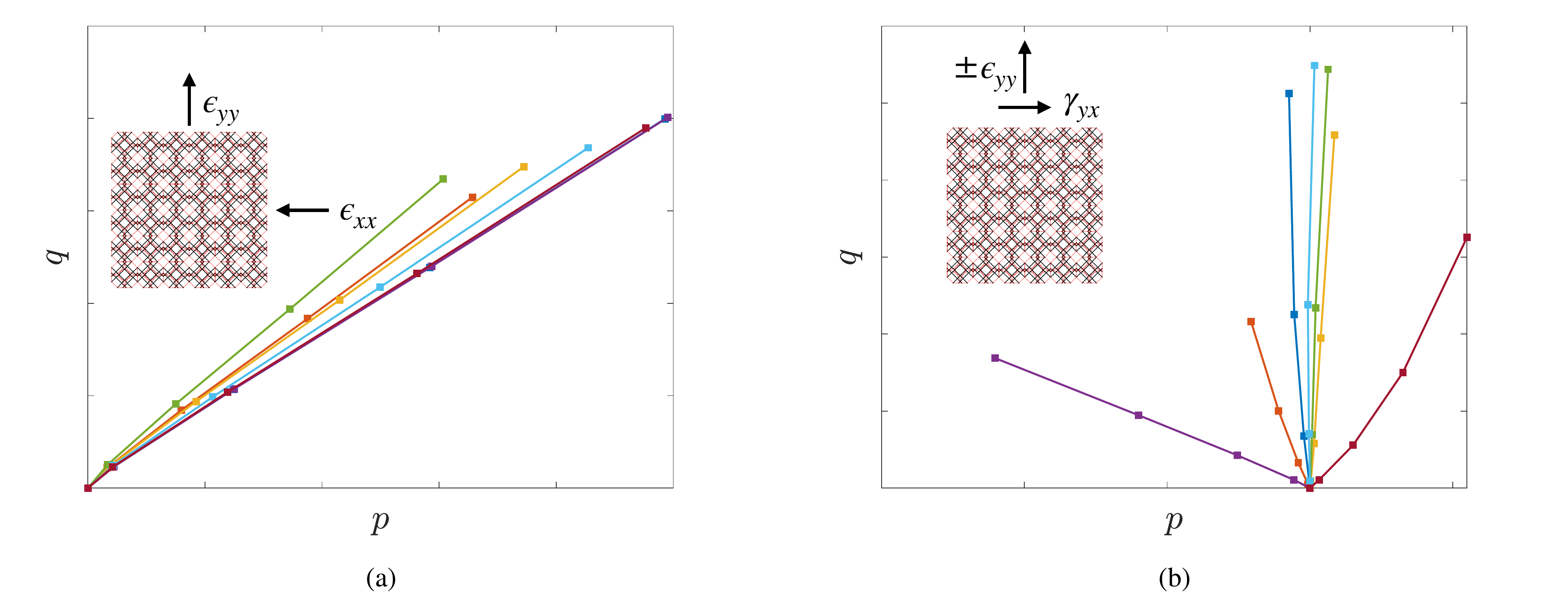} }
 \caption{Schematic of the resulting $p$--$q$ paths of the RVEs. Several loading paths are randomly selected in each case for illustration.}
 \label{fig:data_sample}
\end{figure}

To generate the material dataset $D$, we first determine the size of the 3D representative volume element (RVE) at which energy convergence is observed. We construct 3D models with sizes up to 6$\times$6$\times$1 cells and measure the strain energy vs.~cell size in log-log scale to assess convergence. Based on the result, we select a 3$\times$3$\times$1 3D RVE and perform randomly selected strain-controlled loading paths to generate the dataset from the homogenized response, considering periodic boundary conditions (PBCs) in the $x$ and $y$ directions. For the uniaxial tensile test, we randomly sample $\epsilon_{xx}$ from $[-0.04, 0]$ and $\epsilon_{yy}$ from $[0, 0.08]$. For the simple shear test, we randomly sample $\epsilon_{yy}$ from $[-0.1, 0.1]$ and $\gamma_{yx}$ from $[0, 0.08]$. Several representative loading cases are shown in Fig.~\ref{fig:data_sample} in terms of the resulting $p$--$q$ paths, where $p$ is the hydrostatic pressure,
\begin{equation}
p = -({1}/{3}) \mathrm{tr}(\bm \sigma)\,,
\end{equation}
and $q$ is the effective von Mises stress,
\begin{equation}
q = \sqrt{({3}/{2}) \bm s : \bm s }\,,
\end{equation}
where the deviatoric stress $\bm s$ reads
\begin{equation}
\bm s = \bm \sigma + p\bm I\,,
\end{equation}
with $\bm I$ denoting the second-order identity tensor.

\begin{figure}[t!]
      \centering
      {\includegraphics[clip, trim = 0cm 0cm 20cm 20cm, width=0.75\textwidth]{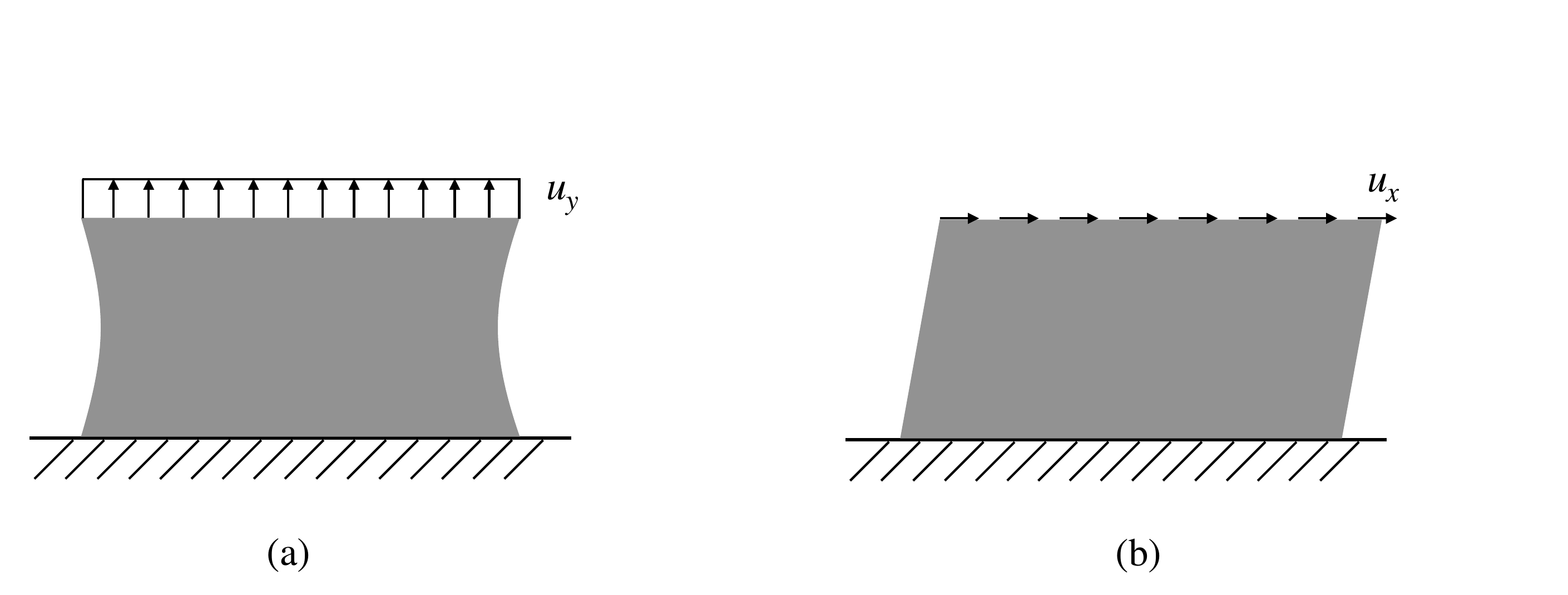} }
     \caption{Schematic of tests performed on the tensegrity monolayer. (a) 5\% strain tension test; (b) 5\% strain shear test.}
 \label{fig:test_schematic}
\end{figure}

In the monolayer uniaxial stretch of 5\% strain (Fig.~\ref{fig:test_schematic}a), we compare the hydrostatic pressure $p$ predicted by the DD framework, considering both the material solution and the mechanical solution, to the actual tensegrity response from the direct numerical simulation. We calculate the hydrostatic stress in the monolayer, and further visualize it with respect to the original nodal positions of the tensegrity cells. 

\begin{figure}[t!]
      \centering
      {\includegraphics[clip, trim = 2cm 0cm 2cm 4cm, width=1\textwidth]{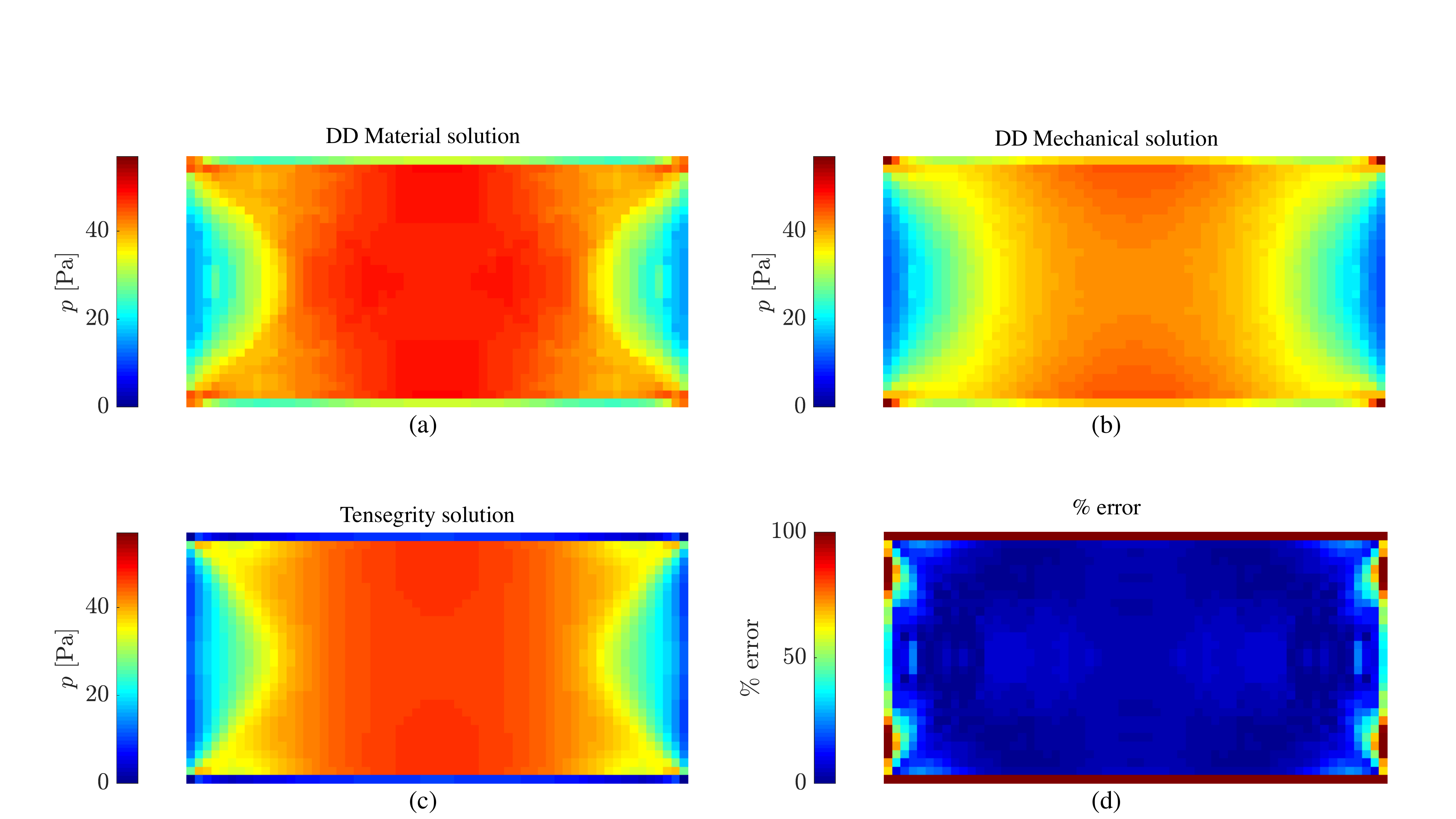} }
     \caption{Hydrostatic pressure field of (a) DD material solution, (b) DD mechanical solution, (c) tensegrity solution, and (d) relative error.}
 \label{fig:stretch}
\end{figure}

Fig.~\ref{fig:stretch} shows a comparison of the DD material solution, the DD mechanical solution, the tensegrity solution, and the percent error map between the DD mechanical solution and the tensegrity solution. We see that the DD solutions qualitatively capture the hydrostatic pressure of the monolayer. From the error map, we observe higher percent errors near the edges. This is primarily due to a mismatch between the local deformation states near the boundaries in the direct tensegrity simulation and those represented in the RVE dataset used for the DD simulations, which were generated under PBCs. The PBCs do not capture the non-periodic deformation patterns present near boundaries. Another factor contributing to the large error near the vertical edges is that the stress magnitude from the tensegrity solution in these regions is small, resulting in a small denominator and thus amplifying the relative error. Aside from the edge regions, the DD solutions show overall good agreement with the direct tensegrity solution.

\begin{figure}[t!]
      \centering
      {\includegraphics[clip, trim = 0cm 0cm 0cm 10cm, width=1\textwidth]{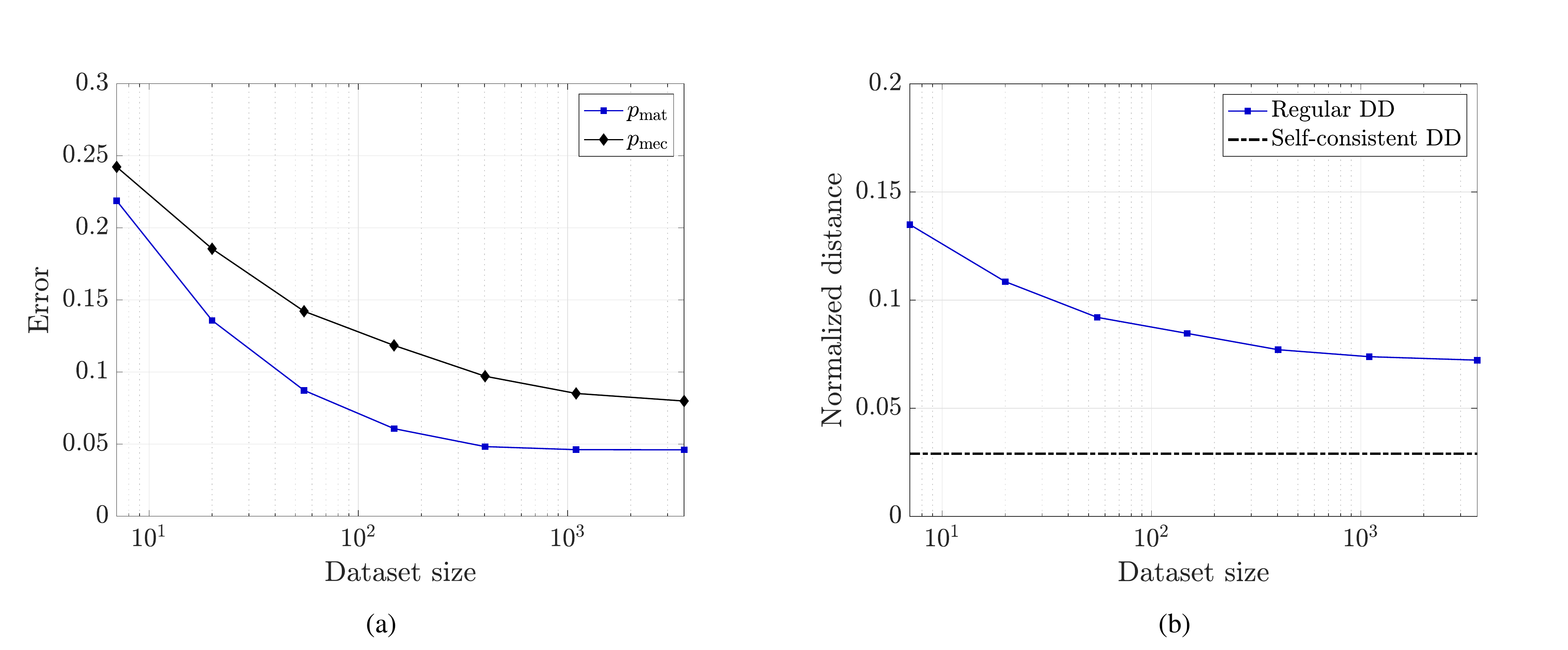} }
     \caption{(a) Relative error between $p_\square$ and $p_\mathrm{tsg}$ vs.~dataset size. The relative error is calculated as $({p_\square - p_\mathrm{tsg}})/{p_\mathrm{tsg}}$. The $\square$ subscript represents either the DD material or mechanical solution. The value is averaged among ten trials. Boundary cells are excluded from the calculation. (b) Normalized distance vs.~dataset size. The normalized distance is calculated as $({\|\mathbf{y}_\mathrm{mat}- \mathbf{y}_\mathrm{mec} \|})/{\|\mathbf{y}_\mathrm{mec}\|}$. The error values are averaged among ten trials. The dotted line shows the benchmark distance calculated using the self-consistent DD simulation.}
 \label{fig:gdist_stretch}
\end{figure}

To assess the dependence of the DD simulations on the number of data points, we plot, for different dataset sizes, the relative error between the hydrostatic pressure from the DD solution and the reference hydrostatic pressure from the direct numerical simulation. Since the deformation states near the outermost boundary cells in the monolayer are not well captured by the RVE dataset generated under PBCs, these cells are excluded from this calculation. To obtain different dataset sizes, we shuffle and randomly sample varying numbers of data points. For each dataset size, we compute the average error over ten trials. In Fig.~\ref{fig:gdist_stretch}a, we observe that the accuracy increases with dataset size, with the final relative error at $\sim$8\% for the mechanical solution and $\sim$5\% for the material solution.

We also calculate the distance between the material solution $\bm y_\mathrm{mat}$ and the mechanical solution $\bm y_\mathrm{mec}$ using Eqns.~\ref{eqn:z} and~\ref{eqn:ze}, normalized with the norm of mechanical solution. We then take the average among ten trials and plot the result for different dataset sizes in Fig.~\ref{fig:gdist_stretch}b. We observe that as the dataset size increases, the normalized distance between the DD solutions decreases to a final value of $\sim$7\%. As a reference, we also plot the normalized distance obtained from a self-consistent DD simulation, where we sample directly from the reference tensegrity response (direct numerical simulation) at the same location. The self-consistent DD simulation yields an error of $\sim$3\%. The general DD solution yields an error value only 4\% higher than this benchmark, highlighting the reliability and limitations of the data sampled from the~RVEs.

\begin{figure}[t!]
      \centering
      {\includegraphics[clip, trim = 2cm 0cm 2cm 4cm, width=1\textwidth]{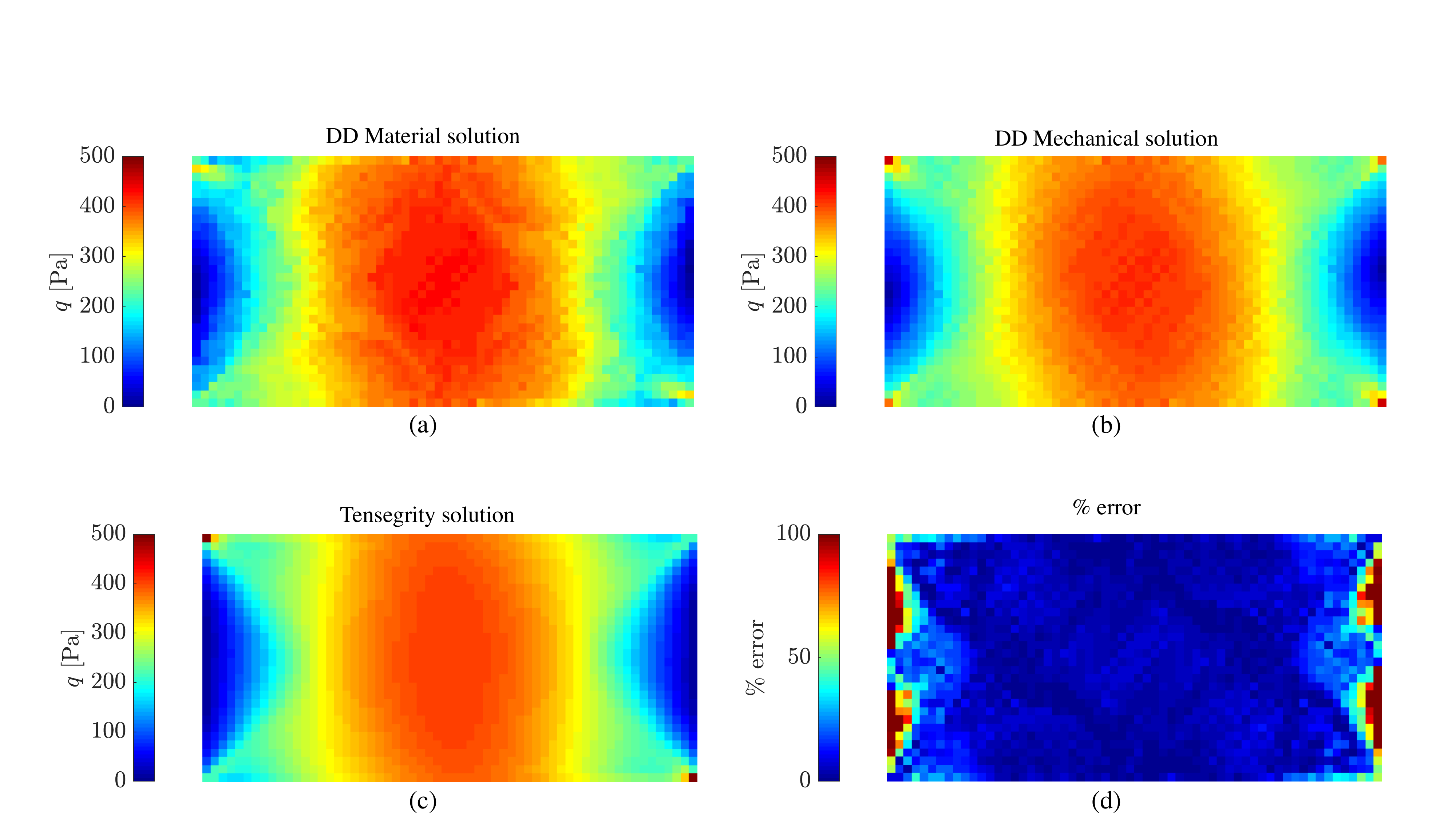} }
     \caption{Deviatoric stress field of (a) DD material solution, (b) DD mechanical solution, (c) tensegrity solution, and (d) relative error.}
 \label{fig:shear}
\end{figure}

Similarly, in the monolayer simple shear test under 5\% strain (Fig.~\ref{fig:test_schematic}b), we compute the average effective stress $q$. The results of the DD simulation are shown in Fig.~\ref{fig:shear} and compared to the reference tensegrity response. Aside from the large error along the edges, primarily due to the mismatch between the local deformation states in those regions and the periodic conditions under which the RVE data is generated, as well as the small stress magnitude near the edges, the DD solutions show good agreement with the tensegrity response obtained from direct numerical simulation.

\begin{figure}[t!]
      \centering
       {\includegraphics[clip, trim = 0cm 0cm 0cm 10cm, width=1\textwidth]{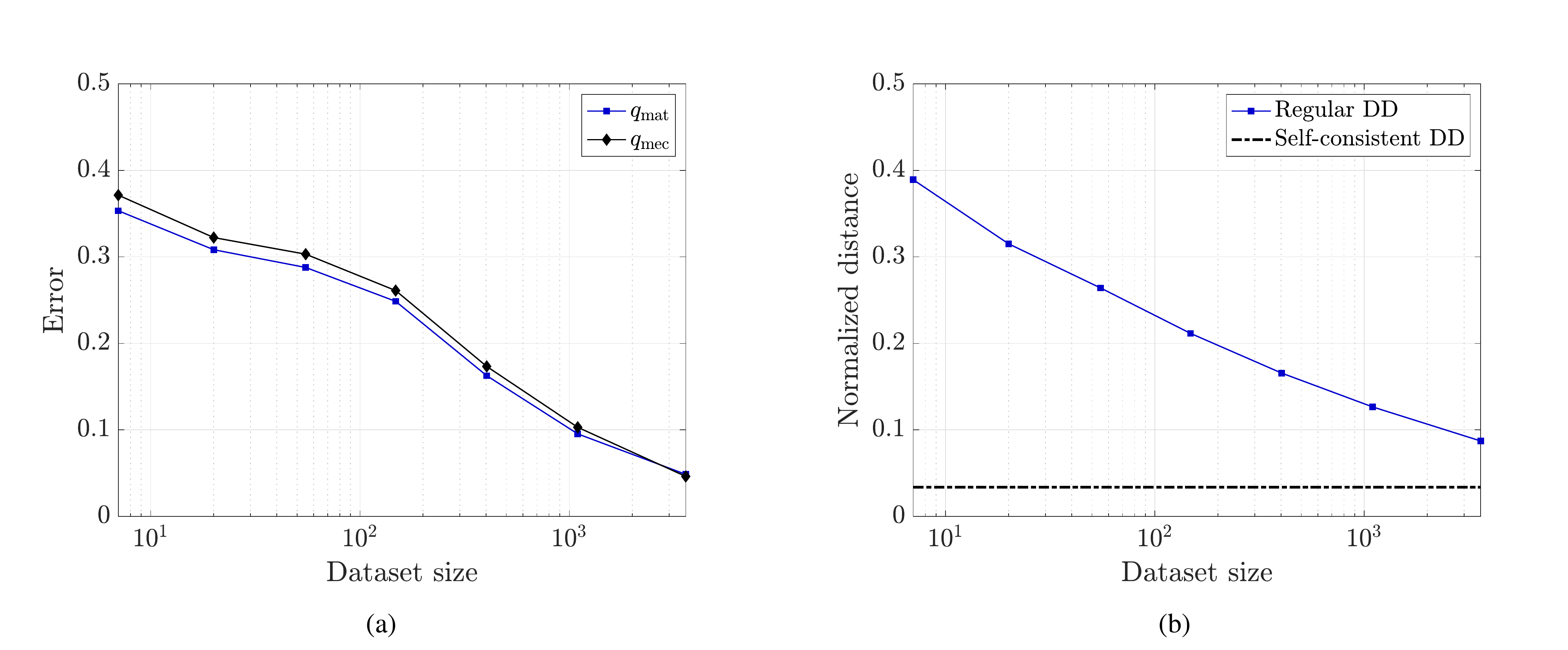} }
        \caption{(a) Averaged relative error between $q_\square$ and $q_\mathrm{tsg}$ vs.~dataset size. (b) Normalized distance vs.~dataset size. The benchmark norm distance calculated from the self-consistent DD simulation is shown in the dotted line.}
 \label{fig:gdist_shear}
\end{figure}

Furthermore, Fig.~\ref{fig:gdist_shear}a shows the relative error between the DD solutions and the direct numerical simulation in terms of $q$, for different dataset sizes, while Fig.~\ref{fig:gdist_shear}b shows the normalized distance between $\bm y_\mathrm{mat}$ and $\bm y_\mathrm{mec}$. Again, we observe that both the relative error and the normalized distance decrease as the dataset size increases. The normalized distance is 5\% higher than the benchmark calculated from the self-consistent DD simulation, indicating that the material data sampled from the RVEs covers most of the phase-space region spanned by the tensegrity structure, although less accurately than in the stretch test.
 

\subsection{Discussion}
The accuracy of the DD algorithm depends on how well the sampled material data spans the desired phase-space region. We suspect several reasons for the observed differences between the DD mechanical solution, the DD material solution, and the direct tensegrity response.

First, we consider the potential sources of error arising from the mismatch between the local states near the boundaries and the material dataset used in the DD framework. In our model, the material dataset is generated from RVEs subjected to PBCs, which capture representative responses of interior regions but do not adequately reflect the deformation states that occur near the boundaries of the tensegrity monolayer. Although we exclude the outermost boundary cells when computing the errors, the boundary effect cannot be entirely eliminated. More sophisticated data sampling strategies, such as data-driven identification methods \cite{stainier2019model}, could help address this limitation in future work.

Second, we suspect that non-simple mechanisms, such as those described by micromorphic or micropolar theories~\cite{Eringen:1964}, are present in the proposed tensegrity structure. As the structure deforms, the angle of the twisted side faces varies, which cannot be captured by the 3D, isotropic, Cauchy-based data-driven framework. In fact, in Fig.~\ref{fig:gdist_stretch}b and Fig.~\ref{fig:gdist_shear}b, we observe around a 3\% normalized distance between $\bm y_\mathrm{mat}$ and $\bm y_\mathrm{mec}$ in the stretch and shear tests using self-consistent DD simulation. This indicates that even when sampling directly from the actual tensegrity response, the material solution still presents a 3\% difference from the mechanical solution based on an equilibrated Cauchy stress. This result suggests that the proposed tensegrity structure may exhibit nontrivial kinematics and kinetics, not captured by standard Cauchy continua. Recent extensions of the DD framework to generalized continua~\cite{karapiperis2021b,ULLOA2024b} could be applied in future work.

Finally, it bears emphasis that the DD framework is implemented in the small-strain setting. Extensions to finite strains~\cite{platzer2021finite} thus emerge as a viable route toward further error reduction.

We acknowledge that for the simple constitutive behavior and small-strain setting considered here, a classical or empirical constitutive model could be more computationally efficient. However, the DD approach provides a flexible framework for future extensions. It enables a parameter-free means of linking microstructure directly to the continuum response, without the need for explicit constitutive laws. It also provides a built-in measure of how well the dataset represents the local mechanical states in a given loading state, which is difficult to assess in classical modeling approaches where error is typically evaluated post hoc against experiments. As richer datasets from experiments or high-fidelity simulations become available, DD frameworks can be leveraged to model more complex or nonlinear behaviors without additional constitutive assumptions.

\FloatBarrier

\section{Conclusions}
\label{sec_conc}
In this paper, we proposed a 3D multicellular tensegrity structure for simulating cell mechanics. We demonstrated that this model can capture the response of single cells and multicellular monolayers. Moreover, we described the 3D heterogeneous stress distribution within an MCS by assigning non-uniform initial prestress to the tensegrity building blocks. This capability positions the model as a potential tool for studying more advanced cellular processes, such as mechanotransduction \cite{Hu2003,Norrelykke2002}. The synergistic interplay of the CSK is indispensable for cellular functionalities. The ability of the tensegrity model to link mechanical forces to specific load-bearing members in 3D can potentially shed new light on the CSK rearrangements and their effects on cellular physiological functions, potentially bridging an important step toward understanding tissue mechanics and its fundamental role in health and disease \cite{Bashirzadeh2019,Efremov2021}.

The tensegrity design and material law adopted in this study are simplified representations aimed at demonstrating the feasibility of the proposed approach. Specifically, the use of a fixed unit geometry and a linear stress-strain constitutive relation\thinspace---\thinspace{albeit} incorporating geometric nonlinearity\thinspace---\thinspace{does} not fully capture the complexity of real CSK components. These modeling choices limit the model’s ability to reflect nonlinear material behavior. Future work will focus on exploring a broader range of tensegrity architectures and more realistic, potentially nonlinear, deformation-dependent material laws, as well as rate-dependent and dynamic conditions.

We also adopted a DD framework for continuum cell mechanics simulations in small deformations. Overall, DD solutions are in good agreement with the tensegrity solution. In general, the material dataset $D$ can be populated using lower-scale simulations, as shown in Section \ref{sec:data_sampling}, or even extracted from existing numerical solutions or experimental data. In this way, the DD framework bypasses the need for a parameter-based continuum model, directly formulating the macroscopic cell mechanics problem based on material data. Combined with homogenization techniques, DD computing may yield accurate results from multiscale analysis while ensuring computational efficiency. This DD-based framework could pave the path for studying the mechanobiology of large-scale cellular structures, such as organs. However, given the mismatch between boundary-localized deformation states and datasets generated under periodic conditions, as well as non-simple mechanical behavior suspected in tensegrity structures, better data sampling methods and extensions to generalized continua and/or non-linear kinematics offer promising directions for future work.

\FloatBarrier

\section*{Declarations}
\subsection*{Acknowledgement}
The authors thank the support from the National Science Foundation (NSF) under award number CMMI-2033779, and the U.S. Army Research Office under grant number W911NF-19-1-0245.

\subsection*{Competing interest}
The authors have no relevant financial or non-financial interests to disclose.


\small
\bibliography{literature}

\end{document}